\documentclass[a4paper,11pt]{article}
\pdfoutput=1 % if your are submitting a pdflatex (i.e. if you have
\usepackage{jcappub} % for details on the use of the package, please
\usepackage[T1]{fontenc} % if needed
\usepackage{aas_macros} %Journal Abbreviations used in the ADS BibTeX
\usepackage{multirow}
\usepackage{xcolor}

\title{Constraining the spatial curvature with cosmic expansion history in a cosmological model with a non-standard sound horizon}

\author[a]{Jordan Stevens,}
\author[a]{Hasti Khoraminezhad,}
\author[a,b]{and Shun Saito}

\affiliation[a]{Institute for Multi-messenger Astrophysics and Cosmology, Department of Physics,\\
Missouri University of Science and Technology, 
1315 N. Pine St., Rolla MO 65409, USA}
\affiliation[b]{Kavli Institute for the Physics and Mathematics of the Universe (WPI),\\ 
Todai Institutes for Advanced Study, the University of Tokyo, Kashiwanoha, Kashiwa, Chiba 277-8583, Japan}

\emailAdd{jbscvp@mst.edu}
\emailAdd{hk3b3@mst.edu}
\emailAdd{saitos@mst.edu}

\date{}
\usepackage{etoolbox}
\makeatletter
\patchcmd{\maketitle}
  {\the\auth@toks\par}
  {\the\auth@toks\par\textmd{\@date}\par}
  {}{}
\makeatother

\abstract{
Spatial curvature is one of the most fundamental parameters in our current concordance flat $\Lambda$CDM model of the Universe. 
The goal of this work is to investigate how the constraint on the spatial curvature is affected by an assumption on the sound horizon scale. 
The sound horizon is an essential quantity to use the standard ruler from the Cosmic Microwave Background (CMB) and Baryon Acoustic Oscillations (BAOs). 
As an example, we study the curvature constraint in an axion-like Early Dark Energy (EDE) model in light of recent cosmological datasets from Planck, the South Pole Telescope (SPT), and the Atacama Cosmology Telescope (ACT), as well as BAO data compiled in Sloan Digital Sky Survey Data Release 16. 
We find that, independent of the CMB datasets, the EDE model parameters are constrained only by the CMB power spectra as precisely and consistently as the flat case in previous work, even with the spatial curvature. We also demonstrate that combining CMB with BAO is extremely powerful to constrain the curvature parameter even with a reduction of the sound-horizon scale in an EDE model, resulting in $\Omega_K=-0.0058\pm 0.0031$ in the case of ACT+BAO after marginalizing over the parameters of the EDE model. 
This constraint is as competitive as the Planck+BAO result in a $\Lambda$CDM model, $\Omega_{K}=-0.0001\pm 0.0018$. 
}
\begin{document}
\maketitle
\flushbottom

%%%%%%%%%%%%%%%%%%%%%%%%%%%%%%%%%%%%%%%%%%%%%%%%%%%%%%%%%%%%%%%%%%%%%%%%%%%%%%%%%%%%%
%%%%%%%%%%%%%%%%%%%%%%%%%%%%%%%%%%%%%%%%%%%%%%%%%%%%%%%%%%%%%%%%%%%%%%%%%%%%%%%%%%%%%
\section{Introduction}
\label{sec:intro}

The spatial curvature of the Universe is one of the most fundamental physical parameters, and spatial flatness is an essential ingredient of the current concordance $\Lambda$CDM cosmological model \cite{Planck2020}.
This work is driven by simple questions; why do we believe that the existing observational datasets suggest the spatial flatness? 
To what extent does this question depend on the assumptions we make in a standard cosmological scenario? 
These questions are partly motivated by the Hubble tension \cite{Riess:2021aa} which offers us excellent opportunities not only to reveal new physics but also to challenge any assumptions behind a concordance flat  $\Lambda$CDM model. 
Many possible solutions to the Hubble tension have been proposed (see \cite{Di_Valentino_2021} for a review). 
Not surprisingly, the vast majority of such studies minimally extended the concordance flat $\Lambda$CDM model (see e.g. \cite{Poulin:2018bb,Fondi:2022tt} for exceptions) so that an extended model is still based on the success of the concordance model. 
Nevertheless, it is not trivial whether and how a minimal extension best explains a series of cosmological observations. \par

\begin{figure}[t]
    \centering
    \includegraphics [width=0.99\textwidth]{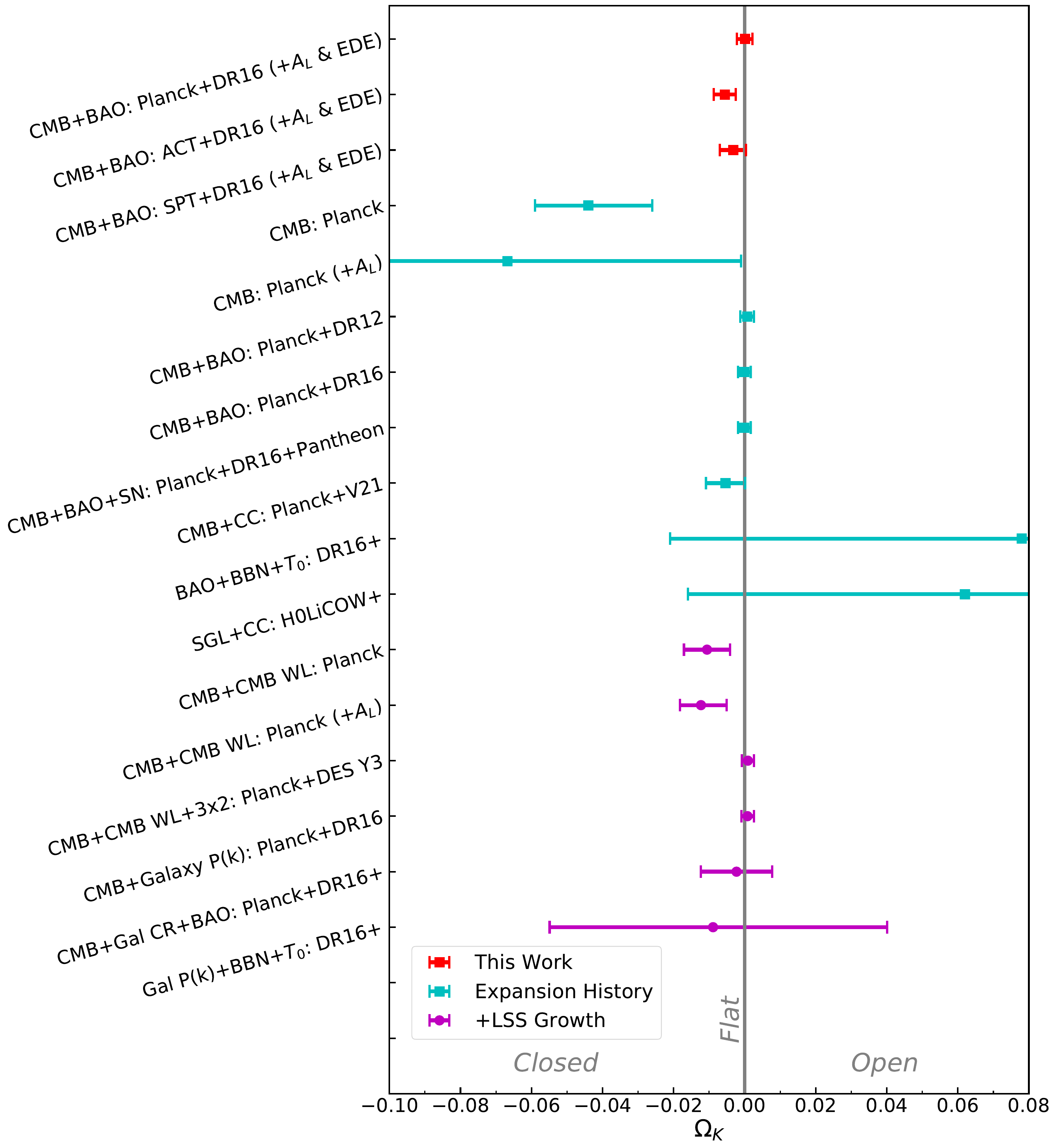}
    \caption{A summary (but highly incomplete examples) of recent measurements (mean with upper and lower 68\% confidence level (C.L.)) of the spatial curvature parameter $\Omega_{K}$ after Planck 2018. 
    We show our main results in this work (red, see Sec.~\ref{sec:results}), the measurements from the observables of the cosmic expansion history (cyan), and the measurements combined with information from the large-scale structure (LSS, magenta).
    The measurements include Planck \cite{Planck2020}, Planck after marginalized over $A_{L}$ \cite{DiValentino:2020na}, Planck + Sloan Digital Sky Survey (SDSS) Data Release 12 (DR12)  \cite{Alam:2017mn}, Planck + SDSS DR16, Planck + SDSS DR16 + Pantheon, 
    Planck + cosmic chornometer (CC) compiled in \cite{Vagnozzi:2021ap} (V21), 
    DR16 with a Big Bang Nucleosynthesis prior \cite{Alam:2021eb}, strong gravitational lensing (SGL) + CC \cite{Liu:2022cc} (see also \cite{Kumar:2021sg,Dhawan:2021cc}), Planck + Dark Energy Survey Year 3 (DES Y3) \cite{DESY3:2022ex}, Planck + the full shape analysis of the galaxy power spectra ($P(k)$) \cite{Simon:2022ar} (see also \cite{Vagnozzi:2021pd}), the full shape analysis of the galaxy $P(k)$ with a BBN prior \cite{Glanville:2022mn}, and Planck + the galaxy clustering ratio (CR) \cite{Bel:2022cr}. 
    }
    \label{Figure:OmKcomp}
\end{figure}

Let us first summarize how spatial curvature is constrained within the context of a $\Lambda$CDM model. 
In Fig.~\ref{Figure:OmKcomp}, we show a summary of examples from recent measurements. 
The spatial curvature is commonly parameterized by a dimensionless parameter, $\Omega_{K}\equiv -c^{2}K/H_{0}^{2}$ where $c$ is the speed of light, $H_{0}=100h\,[{\rm km/s/Mpc}]$ is the Hubble constant, and $K$ is the curvature parameter. 
$\Omega_{K}>0\,(K<0)$, $\Omega_{K}=0\,(K=0)$, and $\Omega_{K}<0\,(K>0)$ correspond to an open, flat, and closed universe, respectively. 
The spatial curvature cannot be determined solely by the information from the primary anisotropies of the cosmic microwave background (CMB, hereafter) due to geometric degeneracy. 
As discussed in detail in \cite{Planck2020}, the CMB temperature and polarization data in Planck preferred the closed universe ($\Omega_{K}=-0.044^{+0.018}_{-0.015}$ for TT,TE,EE+lowE) at $\sim 3\sigma$ level. 
This negative curvature was driven by a smooth temperature power spectrum at $\ell\gtrsim 1000$, which essentially degenerates with the effect of the lensing amplitude, $A_{L}>1$ \cite{Planck:2015,DiValentino:2020na}. 
The preference of the closed universe becomes less significant when the Planck temperature and polarization data was combined with the lensing reconstruction ($\Omega_{K}=-0.0106\pm 0.0065$) and the Baryon Acoustic Oscillation (BAO) data ($\Omega_{K}=0.0007\pm 0.0019$) \cite{Planck2020}. 
In addition, ref.~\cite{Alam:2021eb} showed that $\Omega_{K}=0.078^{+0.086}_{-0.099}$ was obtained only from a recent compilation of the BAO measurements from the Sloan Digital Sky Survey, combined with a prior on the baryon density, $\Omega_{\rm b}h^{2}$ (and hence on the sound horizon scale) from Big Bang Nucleosynthesis and CMB monopole temperature from the COBE/FIRAS data.
%=========================%
%=========================%
Adding the Planck temperature and polarization data (without marginalizing over $A_{L}$) to the compiled BAO data gave $\Omega_{K}=-0.0001\pm 0.0018$, which had no change with the Pantheon Type Ia supernovae \cite{Alam:2021eb} (see also \cite{Zhang:2023ar}). 
Note that including the curvature parameter in a $\Lambda$CDM model does not help mitigating the Hubble tension, since a positive curvature density or open universe is required to reduce the distance to the CMB last scattering with the angular scale of the sound horizon scale being fixed. 
It is important to realize that these results are based on a $\Lambda$CDM model and a standard thermal history to predict the sound horizon scale (see, e.g., \cite{Zhai:2020jc,Yang:2022ex,Semenaite:2022sa} and their references therein for studies in models extended beyond $\Lambda$CDM).
We aim to understand how these results and interpretations are altered when we consider a scenario which alters the sound horizon scale.\par 

As a working example, we consider an axion-like Early Dark Energy (EDE) model \cite{Poulin:2018aa}. 
Among many possible solutions, the EDE model mitigates the Hubble tension by introducing an additional scalar field prior to the recombination epoch and consequently lowering the sound horizon scale \cite{Poulin:2019aa,Smith:2019ih} (see \cite{Lin:2022ar,Galaverni:2023ar,Eskilt:2023ar} for the cosmic birefringence and \cite{Takahashi:2023ar} for the big bang nucleosynthesis). 
The EDE model has recently been given close attention because there is mild evidence of a non-zero EDE component from some observations of the CMB anisotropies. 
Refs.~\cite{Hill:2022pr,Poulin:2021ac,Smith:2022aa} reported evidence at the $\sim 3\sigma$ level from Atacama Cosmology Telescope (ACT) and South Pole Telescope (SPT), while Ref.~\cite{Hill:2020pl,Smith:2022aa} showed little evidence from the Planck data (see also \cite{Handley:2021pd,DiValentino:2023mp} for the discrepancy between the different CMB datasets). 
In addition, there is an attempt to introduce a model with both early- and late-time dark energy fields which could be better fit to CMB, BAO, and $H_{0}$ dataset \cite{Rezazadeh:2022ar}.
It is worth noting that all of these previous work assumed spatial flatness for simplicity (see \cite{Fondi:2022tt,Jiang:2023ar} for an exception and \cite{Hayashi2022de} for including the isocurvature perturbation).\par 

In this paper, we will focus mainly on the observables related to the background expansion through distance measurements, and thus we will not touch upon the large-scale structure (LSS) or the perturbation part (except CMB). 
Information from the CMB lensing is marginalized with $A_{L}$ being varied.
It is well recognized that allowing non-zero EDE is compensated with higher matter density from the fit to CMB data \cite{Hill:2020pl,Smith:2022aa}, which may conflict with the low-lensing (or $S_{8}$) tension (see e.g., \cite{Leauthaud:2017aa,Amon:2022aa}) as shown by refs.~\cite{Ivanov:2020fs,Jedamzik:2021cm,Herold:2022pl,D'Amico:2021aa,Simon:2022ar,Reboucas:2023aa,Goldstein:2023ar,Cruz:2023ar,Ye:2023ar}.
To simultaneously resolve the Hubble and $S_{8}$ tensions, Ref.~\cite{Reeves:2022ss} extended an EDE model by varying the neutrino masses. 
Also, Ref.~\cite{Berghaus:2022aa} considered an EDE scenario where the scalar field decays into extra dark radiation, which could mitigate the $S_{8}$ tension. 
Both of these studies showed that Planck combined with other measurements does \textit{not} prefer an EDE model over the flat $\Lambda$ CDM model. 
Furthermore, Ref.~\cite{Philcox:2022dh} (see also \cite{Smith:2022ra}) measured the distance scale independently of the BAO from the broadband shape (through the matter-radiation equality) of the galaxy power spectrum, reporting a somewhat small Hubble constant, $H_{0}=64.8^{+2.2}_{-2.5}\,{\rm km\,s^{-1}\,Mpc^{-1}}$.  
Again, we notice that these studies assume spatial flatness.
We show some examples of the $\Omega_{K}$ constraints with LSS probes in Fig.~\ref{Figure:OmKcomp}.\par  

The structure of this paper is as follows. 
In Sec.~\ref{sec:theory}, we outline a theoretical background of the observables and physics relevant to this paper. 
It is followed by Sec.~\ref{sec:method} which describes the method and datasets. 
In Sec.~\ref{sec:results}, we report our results, and give elaborated discussion and summary in Sec.~\ref{sec:summary}.  

%%%%%%%%%%%%%%%%%%%%%%%%%%%%%%%%%%%%%%%%%%%%%%%%%%%%%%%%%%%%%%%%%%%%%%%%%%%%%%%%%%%%%
%%%%%%%%%%%%%%%%%%%%%%%%%%%%%%%%%%%%%%%%%%%%%%%%%%%%%%%%%%%%%%%%%%%%%%%%%%%%%%%%%%%%%
\section{Theoretical Background}
\label{sec:theory}

In this section, we briefly provide a theoretical background on the observables relevant to this paper. 
The sound horizon is the comoving distance that a sound wave of the primordial plasma can travel from the beginning of the universe to the point of the last scattering surface of CMB photons, defined by
\begin{equation}
r_{s}(z_{*})=\int_{z_*}^{\infty} \frac{dz}{H(z)}c_{s}(z), 
\label{eq:r_s}
\end{equation}
where $z=1/a-1$ is the cosmological redshift, $a$ is the scale factor of the universe, and $z_{*}\sim 1090$ is the redshift of the CMB last scattering.
The sound speed, $c_{s}(z)$, is given by $c_{s}(z)^{2}=c^{2}/3\{1+R(z)\}$ where $c$ is the speed of light and $R(z)=3\rho_{\rm b}(z)/\{4\rho_{\gamma}(z)\}$ is the baryon-to-photon ratio.
The wave propagation is affected by cosmic expansion through the factor of the Hubble expansion rate, $H(z)\equiv \dot{a}/a$. 
The CMB temperature and polarization anisotropies allow us to infer the angular size of the sound horizon scale, that is, 
\begin{equation}
    \theta_{s} \equiv \frac{r_s(z_*)}{D_A(z_*)}. 
    \label{eq:2.3}
\end{equation}
Here $D_A(z_*)$ is the angular diameter distance at $z=z_{*}$, given by 
\begin{equation}
    D_A(z) = \frac{1}{1+z}S_{K}\left[\int^{z}_{0}\frac{cdz'}{H(z')}\right], 
\end{equation}
where the comoving distance $S_{K}(x)$ is obtained as $\sinh(\sqrt{-K}x)/\sqrt{-K}$ for $K<0$, $x$ for $K=0$, or $\sin(\sqrt{K}x)/\sqrt{K}$ for $K>0$. Notice that $D_{A}(z)\propto H_{0}^{-1}$, independently of the value of $K$. 
A flat $\Lambda$CDM universe provides an excellent fit with the Planck CMB data with $100\theta_{s}=1.04110\pm 0.00031\,{\rm radian}$ and $r_{s}(z_{*})=144.43\pm 0.26\,{\rm Mpc}$ \cite{Planck2020}. 
This leads to $H_{0}=67.36\pm 0.54\,{\rm km/s/Mpc}$ with $\Omega_{\rm m}=0.3153\pm 0.0073$ \cite{Planck2020} which is inconsistent with the local measurement of the Hubble constant, $H_{0}=73.30\pm 1.04\,{\rm km/s/Mpc}$ \cite{Riess:2022ap}.

Baryon Acoustic Oscillations (BAOs) can be precisely measured with a galaxy survey as another standard ruler to measure cosmic expansion at low redshift $z\lesssim 3$. 
Since we measure the BAO scale from a three-dimensional galaxy map, a BAO survey allows us to simultaneously measure 
\begin{equation}
    \theta_{\rm BAO} = \frac{r_{s}(z_{d})}{D_A(z_{\rm BAO})}\,\,\,{\rm and}\,\,\,
    c\Delta z_{\rm BAO} = r_{s}(z_{d})H(z_{\rm BAO}), 
    \label{theta_bao}
\end{equation}
where $z_{\rm BAO}$ is the typical redshift of a BAO survey, and $z_{d}\sim 1060$ is the redshift of the baryon-dragging epoch that occurs slightly after $z_{*}$.  
Planck CMB data provide $r_{s}(z_{d})=147.09\pm 0.26\,{\rm Mpc}$ in a flat $\Lambda$CDM model. 

Since $\theta_{s}$ or $\theta_{\rm BAO}$ is precisely measured with CMB or BAO and is proportional to $r_{s}(z_{*})H_{0}$ or $r_{s}(z_{d})H_{0}$, one way to alleviate the Hubble tension is to reduce the sound horizon scale by introducing new physics prior to the CMB last scattering surface.
A novel Early Dark Energy model exactly achieves this, motivated by an extremely light axion-like scalar field $\phi$ \cite{Poulin:2019aa}.
In general, such a scalar field is modeled with a potential 
\begin{equation}
    V(\phi)=m^{2}\phi^{2} \left\{1-\cos\left(\frac{\phi}{f}\right)\right\}^{n}
    \label{2.5}
\end{equation}
where $f$ is the decay constant of the scalar field.
For the scalar field to be effective before the CMB epoch, the mass $m$ should be smaller than the mass scale corresponding to the Hubble horizon scale at CMB, $m\lesssim 10^{-27}\,{\rm eV}/c^{2}$, so that the field begins to oscillate around the potential minimum. 
Since the potential minimum is locally $V\sim \phi^{2n}$ where the equation of state of the field is given by $w_{\phi}=(n-1)/(n+1)$, we consider $n=3$ such that the cosmic expansion is decelerated with $w_{\phi}<1/3$ (or $n>2$) and hence the sound horizon is reduced. 
The phenomenology of this scalar field can be parameterized by the following effective parameters: $z_c$, critical redshift, which is the redshift at which the EDE contributes to its maximal fraction, $\theta_i \equiv \phi/f$, where $\theta_i$ shows the initial displacement of the scalar field. 
The third parameter is $f_{\rm{EDE}} \equiv {\rm max}(\rho_{\rm{EDE}}(z)/\rho_{\rm tot}(z))$ which indicates the maximal fractional contribution to the total energy density of the universe (see, e.g., \cite{Hill:2020pl}).  
Since $\theta_{s}$ is precisely measured by the frequency of the angular power spectra of the CMB temperature and polarization anisotropies, the EDE parameters are constrained by other physical effects on the CMB spectra. 
The most prominent difference from the $\Lambda$CDM case is the reduction in the diffusion damping scale, which increases the relative power of the CMB spectra at high $\ell$ \cite{Poulin:2018aa}. 

\begin{figure}[t]
    \centering
    \includegraphics [width=1\textwidth]{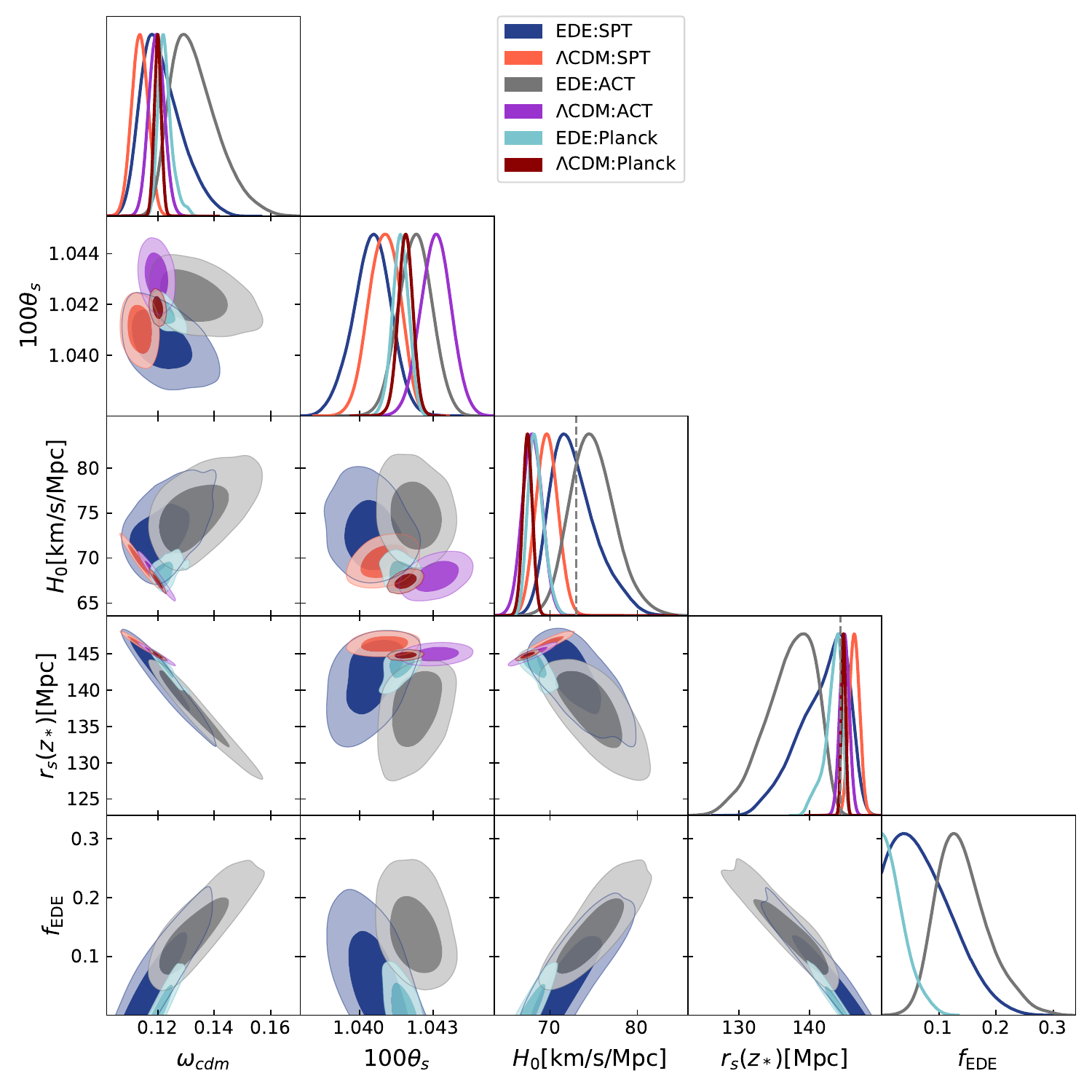}
    \caption{1D and 2D posterior distributions (68\% and 95\% C.L. for the 2D contours) of various cosmological parameters in a flat universe for both $\Lambda$CDM and EDE models. 
    Here we only use the CMB datasets, Planck (cyan/brown for EDE/$\Lambda$CDM), SPT (blue/orange), and ACT (gray/purple). 
    Hereafter, $H_{0}=73.2\,{\rm km/s/Mpc}$ in \cite{Riess:2021aa} and $r_{s}(z_{*})=144.43\,{\rm Mpc}$ in the fidicual flat $\Lambda$CDM fit \cite{Planck2020} are shown as references (vertical dashed lines). 
    }
    \label{Figure:checkflat}
\end{figure}

For our purpose, an EDE model serves as a well-motivated scenario that alters the sound horizon scales, $r_{s}(z_{*})$ and $r_{s}(z_{d})$. In a flat universe, the current dataset allowed $f_{\rm EDE}\lesssim 0.125$ (corresponding to the best-fit value in the ACT case, see Fig.~~\ref{Figure:checkflat}) which reduces both sound horizon scales by a factor of about 5\%. 
Our main goal is to quantitatively study the impact of these reduced sound horizon scales on the constraint on the spatial curvature. 
As a by-product, we provide the constraint on an EDE model, marginalizing over the spatial curvature. 

%%%%%%%%%%%%%%%%%%%%%%%%%%%%%%%%%%%%%%%%%%%%%%%%%%%%%%%%%%%%%%%%%%%%%%%%%%%%%%%%%%%%%
%%%%%%%%%%%%%%%%%%%%%%%%%%%%%%%%%%%%%%%%%%%%%%%%%%%%%%%%%%%%%%%%%%%%%%%%%%%%%%%%%%%%%
\section{Method and Datasets}
\label{sec:method}
We fit the $\Lambda$CDM and EDE models to a series of cosmological data to find the best-fit model parameters and quantify the statistical uncertainties. 
Here, we describe our method and the datasets. 
Since direct sampling of likelihood functions in a high-dimensional parameter space is computationally infeasible, we adopt the Markov chain Monte Carlo (MCMC) method on the basis of Bayesian statistics. 
Note that, instead of sampling posteriors, one could look at the profile likelihood and find confidence intervals using the frequentist approach which helps mitigate projection issues in high-dimensional parameter space \cite{Herold:2022pl,Reeves:2022ss, Herold:2022iib}. 
We perform our series of MCMC sampling using the latest release of publicly available code,  \texttt{MontePython-v3.5}\footnote{\href{https://github.com/brinckmann/montepython\_public}{https://github.com/brinckmann/montepython\_public}} \cite{Audren:2012wb,Brinckmann:2018cvx} interfaced with the \texttt{CLASS\_EDE}\footnote{ \href{https://github.com/mwt5345/class_ede}{https://github.com/mwt5345/class\_ede}} \cite{Hill:2020pl} Boltzmann solver as an extension to the \texttt{CLASS}\footnote{\href{https://github.com/lesgourg/class\_public}{https://github.com/lesgourg/class\_public}} \cite{class-code,CLASS2011} code that accounts for the EDE model. 
We impose flat non-informative priors on both $\Lambda$CDM and EDE parameters, except the prior on $\tau_{\rm reio}$ in the case of SPT-3G and ACT DR4 (see the relevant texts below). 
In the flat $\Lambda$CDM model, we consider the main cosmological parameters $\{ \Omega_{\rm b}h^{2}, \omega_{cdm}\equiv \Omega_{cdm}h^{2}, \theta_s, \ln[10^{10}A_s],n_{s}, \tau_{\rm{reio}} \}$ that account for the amount of baryon and CDM densities in the Universe, the ratio of the sound horizon to the angular diameter distance, the normalization amplitude and the spectral index of the primordial power spectrum, and the reionization optical depth, respectively. 
When we consider a non-flat cosmology, we include two additional parameters, $\{\Omega_{\rm K},A_{L}\}$. 
The reason we add the lensing parameter $A_{L}$ is that there exists a degeneracy between $\Omega_{\rm K}$ and $A_{L}$ in the Planck temperature and polarization data \cite{DiValentino:2020na} (see their Fig.~2), and this helps isolate the lensing information from the dataset we consider (see appendix \ref{sec:appendix} for the results with fixed $A_L=1$). 
Regarding an EDE model, we add three more phenomenological parameters, $\{f_{\rm EDE}, \log_{10}z_{c}, \theta_{i}\}$ using the shooting method described in \cite{Smith:2019ih} to map them to the theoretical parameters $\{f,m\}$ mentioned in Eq.~(\ref{2.5}). 
In modeling free-streaming neutrinos, we take into account the Planck collaboration convention as two massless species and one massive with $M_{\nu}=0.06\, \rm{eV}$. 
In our analysis, we adopt the effective number of relativistic degrees of freedom to its standard model prediction as $N_{\rm eff}=3.046$. 
To compute the non-linear matter power spectrum we use \texttt{Halofit} \cite{Smith:2002dz,Takahashi:2012hf} for CMB lensing, although again the information from CMB lensing is minimized by being marginalized over $A_{L}$. 
The default prior choice is given in Table.~\ref{table:prior}. 
Moreover, depending on the choice of the data set we use, we have additional nuisance parameters. 
For example, for the case of the full Planck 2018 likelihood, we have $47$ nuisance parameters that would be added to our multidimensional parameter space.  
For the case of SPT-3G 2018 likelihood, we have $20$ nuisance parameters. 
In the case of the ACT DR4 likelihood, we only have one nuisance parameter which stands for the polarization efficiency. 
To analyze the chains and produce plots, we use the \texttt{GetDist}\footnote{\href{https://getdist.readthedocs.io/}{https://getdist.readthedocs.io/}} Python package \cite{Lewis:2019xzd}.
We consider chains to be sufficiently converged, checking the Gelman-Rubin criterion $|R-1|\lesssim 0.05$\footnote{Note that, in the case $\rm EDE+ \Omega_K + A_L$ model using SPT combination of dataset without additional BAO dataset we consider the convergence to $|R-1|\lesssim 0.1$ due to the degeneracy of large number of parameters.}. 

\begin{table}[t]
\centering
\renewcommand{\arraystretch}{1.3} % Adjust the value to increase/decrease line spacing
    \begin{tabular}{c c}
    \hline
    Parameters & Prior \\ [0.5ex]
    \hline
       $100\theta_s$ & [0.5,10] \\
       $\rm{ln}10^{10}A_s$ & [1.61,3.96] \\
       $n_s$ & [0.8,1.2] \\
       $\tau_{\rm{reio}}$ & [0.02,\rm{None}] \\
       $\Omega_{K}$  & [-0.5,0.5] \\
       $A_{L}$  & [0.1,2.1] \\
       $\log_{10}z_{c}$ & [3.1,4.3] \\
       $\theta_{i}$ & [0.1,3.1] \\
       $f_{\rm EDE}$ & [0.001,0.5] \\
       \hline
    \end{tabular}
    \caption{The assumed ranges of uniform priors.}
    \label{table:prior}
\end{table}

In this work, we adopt the following datasets:
\begin{description}
    \item[Planck 2018:] We consider the multifrequency TT, TE, and EE power spectra and covariances from Planck PR3 (2018)\footnote{\href{http://pla.esac.esa.int/pla/\#cosmology}{http://pla.esac.esa.int/pla/\#cosmology}} \cite{Planck:2018nkj,Planck2020,Planck:2019nip}, including the publicly available \texttt{Plik\_HM} high-$\ell$ likelihood consisting of $30 \leq \ell \leq 2508$ for TT and $30 \leq \ell \leq 1996$ for TE and EE spectra, and also from the \texttt{Commander} likelihood, we consider low-$\ell$ ($2 \leq \ell \leq 29$) TT data.
    In addition, we include the Planck reconstructed CMB lensing potential power spectrum \cite{Planck:2018lbu}. 
    The gravitational lensing of the CMB has been detected by a high ($40 \sigma$) statistical significance in Planck 2018 \cite{Planck:2018lbu}. The multipole range of the Planck lensing power spectrum covers $8 \leq \ell \leq 400$.
    In summary, we use Planck-high-$\ell$ \texttt{TTTEEE} $+$ Planck-low-$\ell$ \texttt{TT} $+$ Planck-low-$\ell$ \texttt{EE} + Planck-lensing, which we refer to this combination as ``Planck''.
    
    \item[Planck $\ell\leq 650$:] When we use SPT-3G or ACT DR4 instead of Planck 2018, we combine the Planck TT data at $\ell \leq 650$. 
    For this purpose, we use the full \texttt{Plik} likelihood rather than the \texttt{Plik\_lite} version to change the range of the multipole.
    This choice is motivated by the fact that Planck 2018 is fully consistent with WMAP \cite{Hinshaw_2013} up to this multipole \cite{Galli:2017pw} and that our results can be fairly compared with previous SPT and ACT results \cite{Hill:2022pr,Smith:2022aa}.
    We also add a Gaussian prior on the optical depth as $\tau_{\rm reio} = 0.065 \pm 0.015$, following \cite{Hill:2022pr,ACT:2020gnv}. 

    \item[SPT-3G:] We use the most updated SPT-3G data from 2020 data release \cite{SPT-3G:2021eoc} which includes EE and TE power spectra spanning over the angular multipole range, $300 \leq \ell < 3000$, from observations of a $1500\, \rm{deg}^2$ survey field in three frequency bands centered at $95$, $150$, and $220\, \rm{GHz}$\footnote{Note that, while we prepare this work, the SPT-3G TT measurement has been recently released and we do include it in this work\cite{Balkenhol:2022rvc} }. We adopt the publicly available full SPT-3G likelihood \cite{SPT-3G:2021eoc} for the \texttt{MontePython} environment\footnote{\href{https://github.com/ksardase/SPT3G-montepython}{https://github.com/ksardase/SPT3G-montepython}} \cite{Chudaykin:2022rnl}. 
    The likelihood function marginalizes posteriors over the super-smaple lensing parameter, the terms representing Galactic dust emission, polarized dust and the noise from the radio galaxies along with the all-sky calibration of temperature and polarization \cite{SPT-3G:2021eoc}.
    We refer to this SPT-3G with Planck $\ell\leq 650$ (including the $\tau_{\rm reio}$ prior) as ``SPT''.
    
    \item[ACT DR4:] We use the latest version of ACT data from the fourth data release, ACT DR4, \cite{ACT2020} from the 2013-2016 survey covering $>15000\,\, \rm{deg}^2$ including multi-frequency temperature and polarization measurements. 
    The TT, TE and EE power spectra have already been marginalized over various uncertainities such as foreground emission and systematic errors. 
    We use the \texttt{MontePython} support for the publicly available \texttt{actpollite dr4} likelihood implemented in  \texttt{pyactlike}.\footnote{\href{https://github.com/ACTCollaboration/pyactlike}{https://github.com/ACTCollaboration/pyactlike}} 
    The range of multipoles for TE and EE power spectra spans over $326 < \ell < 4325$ while the TT power spectrum covers $ 576 < \ell < 4325 $ \cite{ACT:2020gnv}. 
    The only nuisance parameter in this likelihood is the overall polarization efficiency. 
    We call this ACT DR4 data with Planck $\ell\leq 650$ (including the $\tau_{reio}$ prior) as ``ACT''.
    
    \item[Baryon Acoustic Oscillations (BAOs):] We consider the following datasets from BAO survey as a probe of the cosmic expansion history at low redshifts; 
    the 6df galaxy redshift survey at $z_{\rm BAO}=0.106$ \cite{Beutler_2011},  
    the Sloan Digital Sky Survey (SDSS) Data Release (DR) 7 main galaxy sample at $z_{\rm BAO}=0.15$ \cite{Ross:2014qpa}, and data compiled from the SDSS DR16 Baryon Baryon Oscillation Spectroscopic Survey (BOSS) \cite{Alam:2017mn} and extended BOSS (eBOSS) measurements \cite{deSainteAgathe:2019voe,Alam:2021eb,duMasdesBourboux:2020pck} that include Luminous Red Galaxy sample at $z_{\rm BAO}=0.38$, $0.51$ and $0.698$,\footnote{\href{https://github.com/CobayaSampler/bao_data/blob/master/sdss_DR16_BAOplus_LRG_FSBAO_DMDHfs8.dat}{https://github.com/CobayaSampler/bao\_data/blob/master/sdss\_DR16\_BAOplus\_LRG\_FSBAO\_DMDHfs8 \\ .dat}} the QSO sample at $z=1.48$,\footnote{\href{https://github.com/CobayaSampler/bao_data/blob/master/sdss_DR16_BAOplus_QSO_FSBAO_DMDHfs8.dat}{https://github.com/CobayaSampler/bao\_data/blob/master/sdss\_DR16\_BAOplus\_QSO\_FSBAO\_DMDHfs8 \\ .dat}} (note that for these two aformaentioned cases the covariance matrices have been marginalized over the $f\sigma_8$ measurements), and also the $\rm{Ly}\alpha \times \rm{Ly}\alpha$,\footnote{\href{https://github.com/CobayaSampler/bao_data/blob/master/sdss_DR16_LYAUTO_BAO_DMDHgrid.txt}{https://github.com/CobayaSampler/bao\_data/blob/master/sdss\_DR16\_LYAUTO\_BAO\_DMDHgrid.txt}} and  $\rm{Ly}\alpha \times \rm{QSO}$ measurements at $z=2.334$.\footnote{\href{https://github.com/CobayaSampler/bao_data/blob/master/sdss_DR16_LYxQSO_BAO_DMDHgrid.txt}{https://github.com/CobayaSampler/bao\_data/blob/master/sdss\_DR16\_LYxQSO\_BAO\_DMDHgrid.txt}}
    Note that we do not include the BAO measurement for the eBOSS Emission Line Galaxy as its contribution to the fit is minor. 
    In this work we do not consider the redshift-space distortion or full-shape galaxy power spectrum.
\end{description}

We do not include the distance ladder measurement of the Hubble constant, $H_{0}=73.2\pm 1.4\,{\rm km/s/Mpc}$ \cite{Riess:2021aa} in our fit, but compare it with the inferred values. 
Moreover, since our primary interests are the observables relevant to the sound horizon scale, we do not include the data from Pantheon Type Ia supernovae. 
Previous studies show that the impact of the Pantheon data is generally minor when combined with the BAO data (see e.g., Fig.~\ref{Figure:OmKcomp} and \cite{Yang:2022ex}).

%=========================%
%=========================%

%%%%%%%%%%%%%%%%%%%%%%%%%%%%%%%%%%%%%%%%%%%%%%%%%%%%%%%%%%%%%%%%%%%%%%%%%%%%%%%%%%%%%
%%%%%%%%%%%%%%%%%%%%%%%%%%%%%%%%%%%%%%%%%%%%%%%%%%%%%%%%%%%%%%%%%%%%%%%%%%%%%%%%%%%%%
\section{Results}
\label{sec:results}

We begin with the results of the fit with the CMB datasets only for the flat $\Lambda$CDM and EDE models in Fig.~\ref{Figure:checkflat}. 
The purpose of showing Fig.~\ref{Figure:checkflat} is a sanity check that we have successfully reproduced the previous work; 
Our fits in the flat $\Lambda$CDM are in excellent agreement with the previous work.
Consistently with \cite{SPT-3G:2021eoc}$, \omega_{\rm cdm}$ (and $r_{s}(z_{*})$) in the flat $\Lambda$CDM model for SPT is slightly smaller than that for Planck and ACT. 
In the EDE model, Planck disfavors non-zero $f_{\rm EDE}$, while ACT and SPT somewhat prefer non-zero values, $f_{\rm EDE}< 0.218\,(0.199)$ at $95\%$ C.L. for ACT (SPT). 
Non-zero $f_{\rm EDE}$ in ACT or SPT leads to reduction in the sound horizon scale, $r_{s}(z_{*})$ while $\theta_{s}$ is kept nearly fixed, yielding higher values of $H_{0}$ than flat $\Lambda$CDM cases \cite{Hill:2020pl,Smith:2022aa}.

%=========================%
\begin{figure}[t]
    \centering
    \includegraphics [width=1\textwidth]{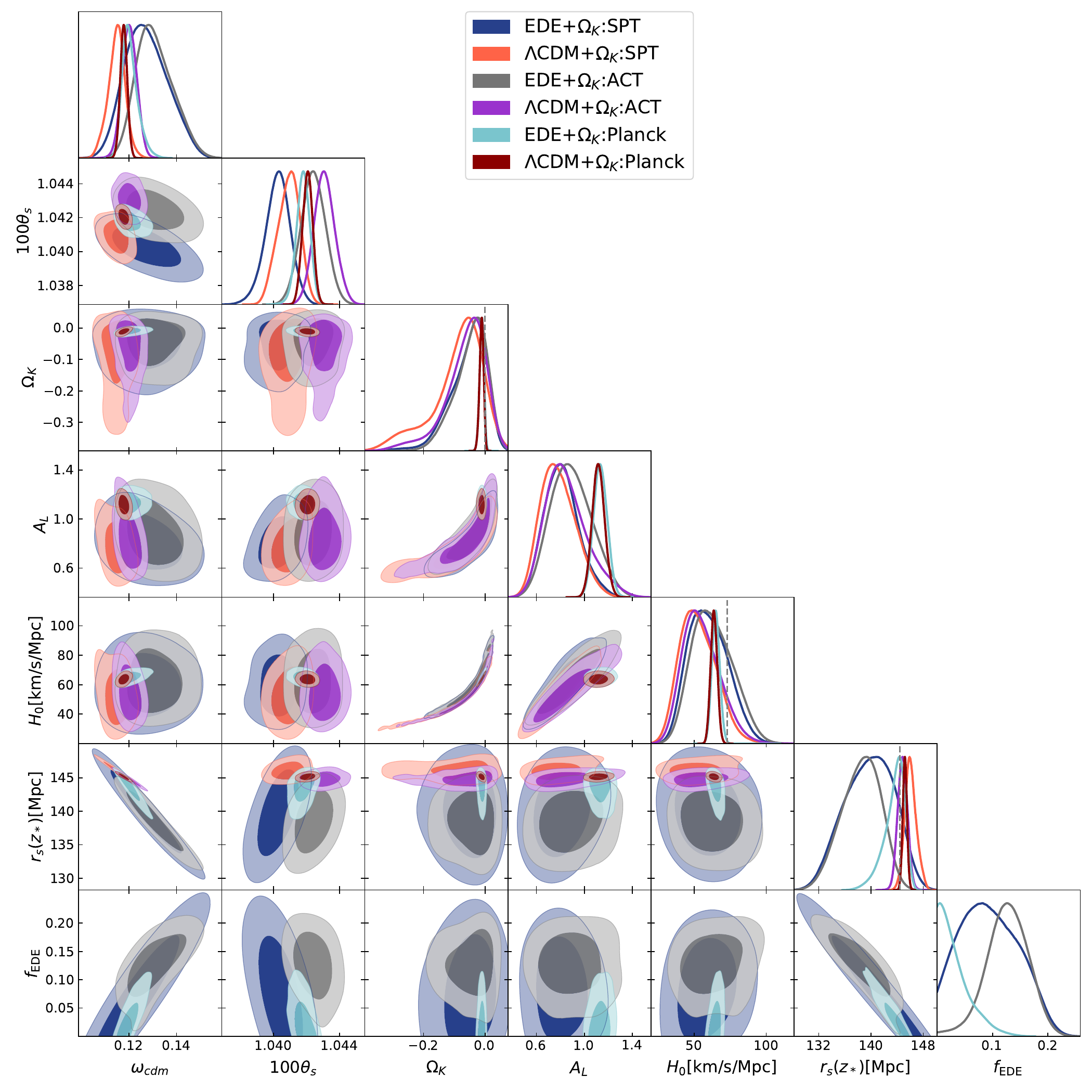}
    \caption{1D and 2D posterior distributions (68\% and 95\% C.L. for the 2D contours) of various cosmological parameters in a curved universe for both $\Lambda$CDM and EDE models. 
    Here we only use the CMB datasets, Planck (cyan/brown for EDE/$\Lambda$CDM), SPT (blue/orange), and ACT (gray/purple). 
    We show $\Omega_{K}=0$ as a reference (vertical dashed line). 
    }
    \label{Figure:curvedcmb}
\end{figure}
%=========================%

\begin{figure}[t]
    \centering
    \includegraphics[width=7.2cm]{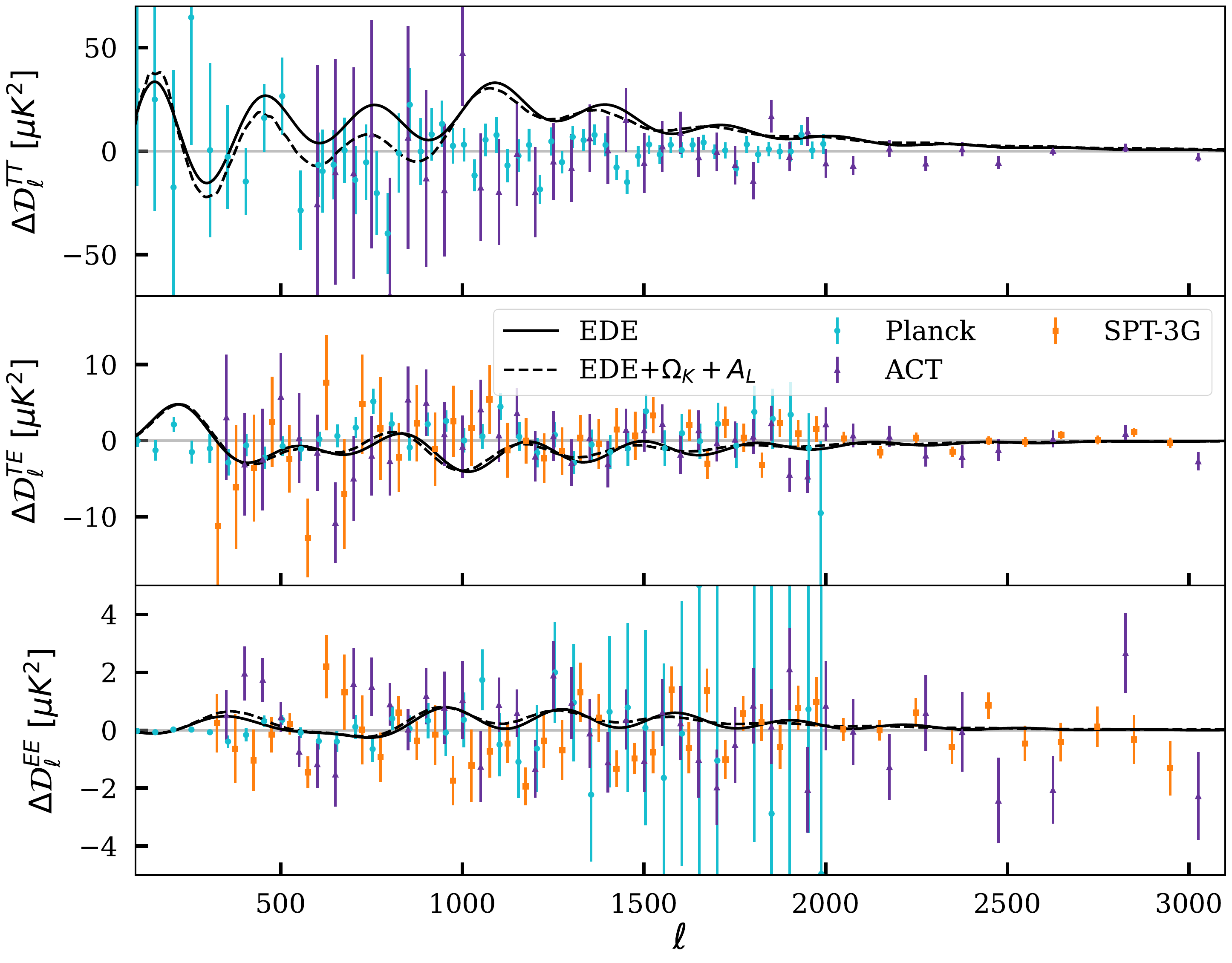}
    \qquad
    \includegraphics[width=7cm]{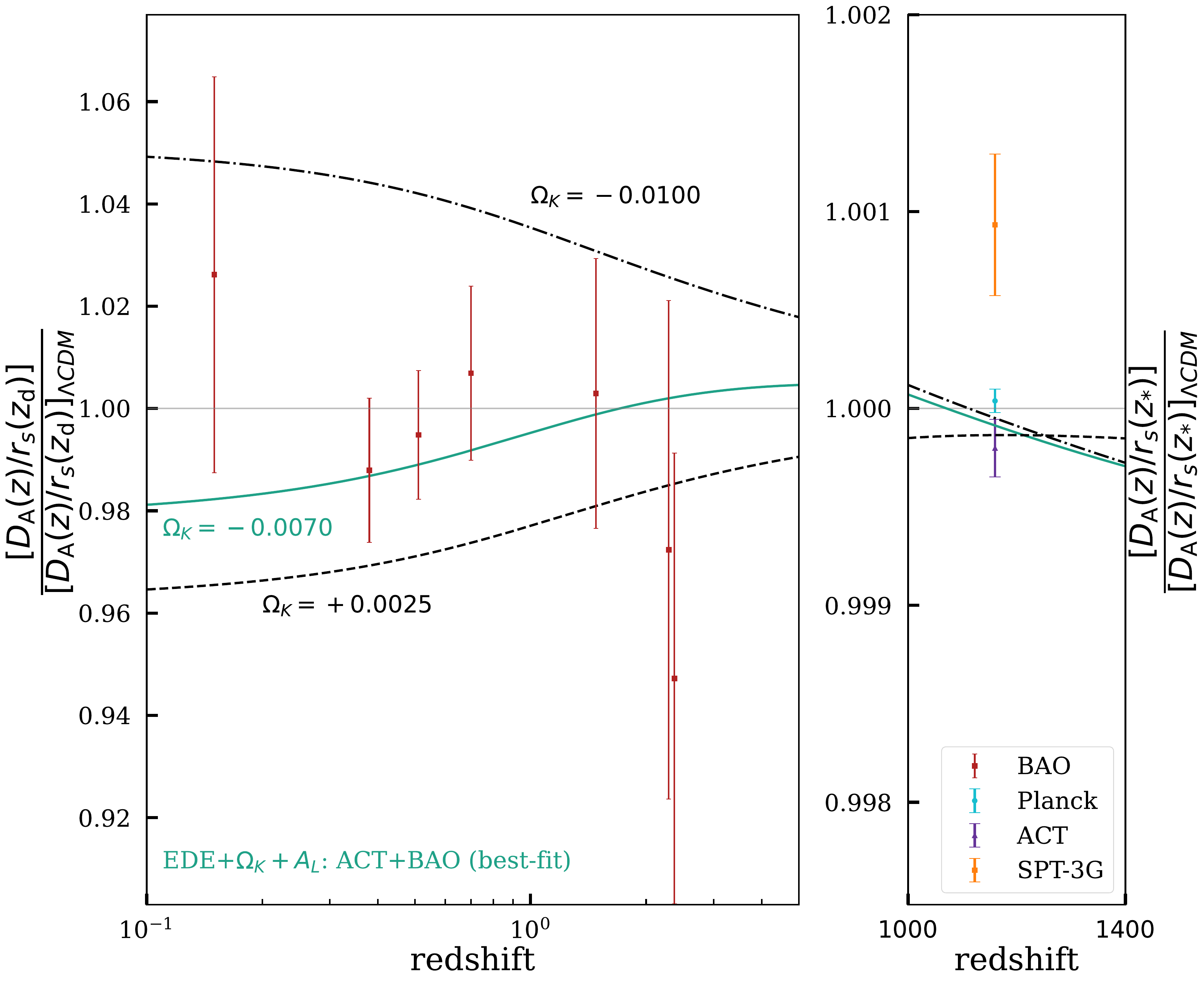}
    \caption{Left Panel: ACT best-fit CMB angular spectra residuals (TT, TE and EE) for EDE (solid) and EDE+$\Omega_K + A_L$ (dashed). Cyan, orange and purple data points show the residuals of the Planck, SPT-3G and ACT measurements with respect to the flat $\Lambda$CDM best-fit in Planck. 
    Right Panel: The ratio of the sound horizon scale (Eqs.~\ref{eq:2.3}, \ref{theta_bao}) measurements in EDE+$\Omega_K + A_L$ model with respect to the Planck flat $\Lambda$CDM. Red data points in the left panel show the low-redshift BAO measurements from DR16 and the right panel shows the CMB measurements. The green curve represents the EDE+$\Omega_K + A_L$ best-fit model in the light of the ACT+BAO datasets.  Black dashed/dotted-dashed curve shows the same model with slightly larger/smaller values for $\Omega_K$.}
    \label{Figure:spectra}
\end{figure}

Next, we present the CMB results including the spatial curvature in Fig.~\ref{Figure:curvedcmb}. The constraints on $f_{\rm EDE}$ with these CMB datasets are not affected by $\Omega_{K}$ and $A_{L}$ and remain similar to the flat case in Fig.~\ref{Figure:checkflat}. 
Meanwhile, we confirm the degeneracy among $\Omega_{K}$, $A_{L}$, and $H_{0}$ as expected, giving weak constraints on these parameters only with the CMB data. 
However, the impact of the EDE model paramaters on the $\Omega_{K}$ constraints is minor \cite{Fondi:2022tt}. 
We have curvature constraints with the CMB datasets even after marginalizing the EDE model parameters; $\Omega_K = -0.0105_{+0.0057}^{+0.0072}$ for Planck, $\Omega_K = -0.0508_{-0.0178}^{+0.0717}$ for ACT, and $\Omega_K = -0.0635_{-0.0225}^{+0.0944}$ for SPT (see also Table \ref{Table:main}).  
The reason why Planck constrains $\Omega_{K}$ better than ACT and SPT is that we combine with the reconstructed CMB lensing in Planck.
Without the Planck CMB lensing, the Planck constraint on $\Omega_{K}$ is degraded from $\Delta \Omega_K \sim 0.0066$ to $\Delta \Omega_K \sim 0.0488$ in the $\Lambda$CDM case (see also \cite{Planck2020}). 
To understand why the EDE model paramaters are constrained even with $\Omega_{K}$, we present the CMB angular power spectra and their residuals from the Planck $\Lambda$CDM best-fit model in the left panel of Fig.~\ref{Figure:spectra}.
Top, middle, and bottom panels show the residuals from TT, TE, and EE spectra, respectively. 
If the data deviates from zero in each panel, a model may be required to include physics beyond the Planck $\Lambda$CDM model. 
Solid and dashed curves show the ACT best-fit curves in EDE and EDE$+\Omega_{K}+A_{L}$ models. 
We only show the ACT best-fit as it prefers the largest $f_{\rm EDE}$. 
As demonstrated by the two curves, including $\Omega_K$ in the EDE model does not affect the acoustic feature of the CMB spectra and consequently has little impact on the EDE parameters.   
This is not surprising, since $\Omega_{K}$ is constrained by the CMB mainly through $\theta_{s}$ with the sound horizon fixed.  
We mitigate information from CMB lensing by marginalizing $A_{L}$.

\begin{table}[t]
\centering
\renewcommand{\arraystretch}{1.7} % Adjust the value to increase/decrease line spacing
\scalebox{0.75}{
\begin{tabular}{c|ccc|ccc}
\hline
 & Planck+BAO& ACT+BAO& SPT+BAO & Planck & ACT & SPT  \\ 
\hline

$f_{\rm EDE}$  & $0.044_{-0.041}^{+0.015}$ & $0.138_{-0.053}^{+0.029}$ & $0.107_{-0.082}^{+0.048}$ & $0.033_{-0.031}^{+0.008}$ & $0.132_{-0.047}^{+0.031}$  & $0.086_{-0.076}^{+0.035}$ \\
$\theta_{i}$ & $2.109_{-0.288}^{+0.942}$ & $1.518_{-1.328}^{+1.443}$ & $1.569_{-1.170}^{+1.144}$ & $1.868_{-1.541}^{+1.149}$ & $1.485_{-1.304}^{+1.468}$ & $1.554_{-0.757}^{+1.033}$\\
$\log_{10}(z_{c})$ & $3.693_{-0.263}^{+0.184}$ &$3.251_{-0.134}^{+0.056}$ & $3.502_{-0.166}^{+0.100}$ & $3.733_{-0.322}^{+0.204}$ & $3.241_{-0.130}^{+0.054}$ & $3.573_{-0.252}^{+0.101}$\\
\hline
 $H_0\,[{\rm km/s/Mpc}]$    & $69.264_{-1.267}^{+0.871}$ & $72.139_{-2.491}^{+1.237}$ & $71.732_{-3.308}^{+1.615}$ & $64.931_{-2.659}^{+2.385}$ & $62.874_{-18.326}^{+12.495}$& $62.113_{-24.602}^{+15.587}$ \\
$\Omega_{K}$  & $-0.0009_{-0.0019}^{+0.0019}$ & $-0.0058_{-0.0031}^{+0.0031}$ &$-0.0032_{-0.0032}^{+0.0032}$ & $-0.0105_{+0.0057}^{+0.0072}$ & $-0.0508_{-0.0178}^{+0.0717}$ &  $-0.0635_{-0.0225}^{+0.0944}$ \\
$r_s(z_{*})\,[{\rm Mpc}]$ & $142.786_{-1.021}^{+2.172}$ & $138.198_{-2.040}^{+4.584}$ & $139.459_{-3.093}^{+6.467}$ & $143.540_{-0.625}^{+1.717}$ & $138.3_{-1.891}^{+4.319}$ & $140.611_{-2.68}^{+5.474}$ \\
\hline 
 \end{tabular}
 }
 \caption{The constraints on key cosmological parameters for the curved-EDE (EDE$+\Omega_{K}+A_{L}$) model. Each column corresponds to different dataset. Each value quoted as $\rm mean_{-(mean - lower \,\,68\% limit)}^{+(upper \,\,68\% limit - mean)}$ where ``mean'' refers to the mean value of the marginalized posterior distribution. }
 \label{Table:main}
\end{table}

%=========================%
%=========================%

%=========================%
\begin{figure}[t]
    \centering
    \includegraphics [width=1\textwidth]{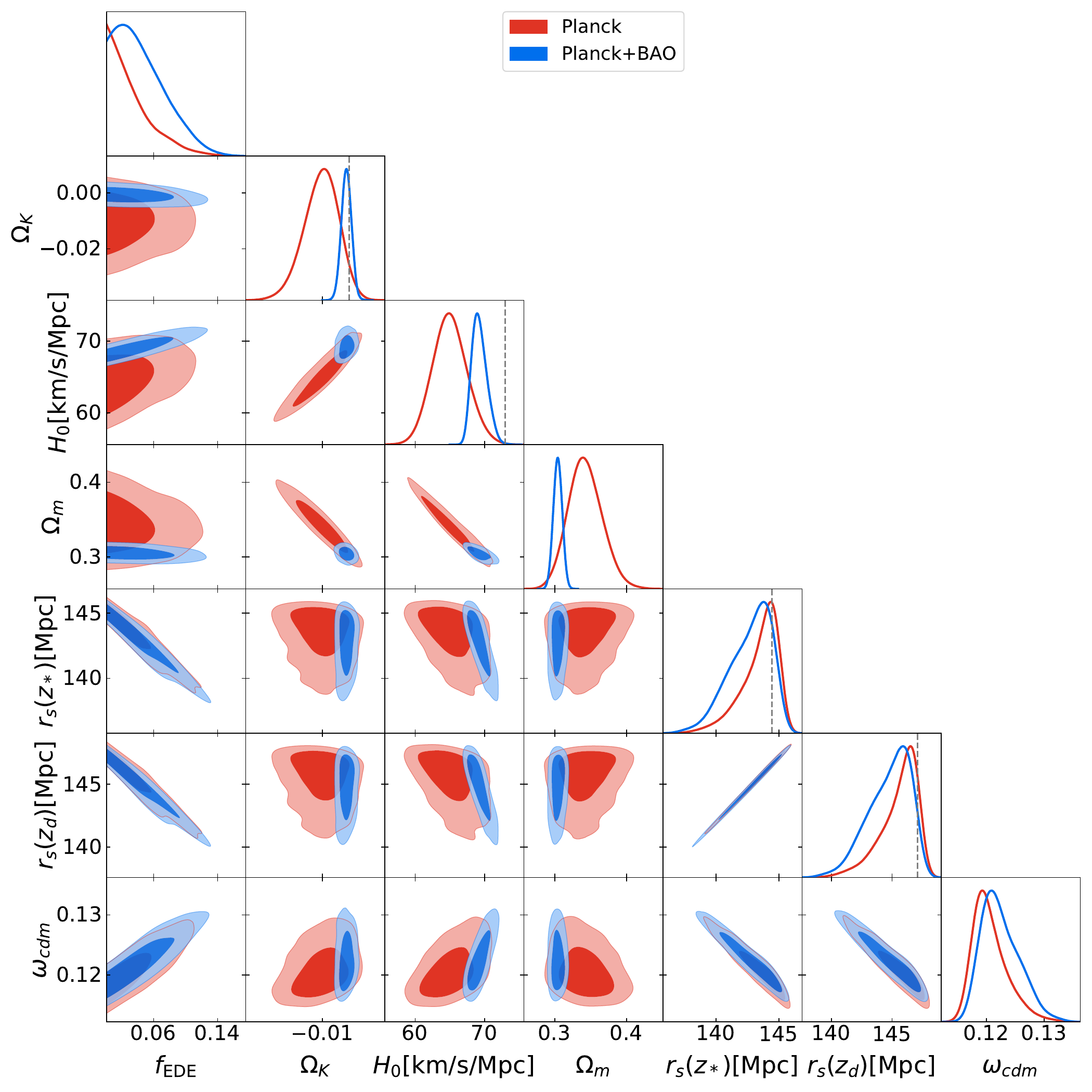}
    \caption{1D and 2D posterior distributions (68\% and 95\% C.L. for the 2D contours) in a curved EDE model with Planck only (red) and Planck+BAO (blue). We show $r_{s}(z_{d})=147.09\,{\rm Mpc}$ in the flat $\Lambda$CDM fit in Planck as a reference (vertical dashed line). 
    }
    \label{Figure:PlanckBAO}
\end{figure}
%=========================%

%=========================%
\begin{figure}[t]
    \centering
    \includegraphics [width=1\textwidth]{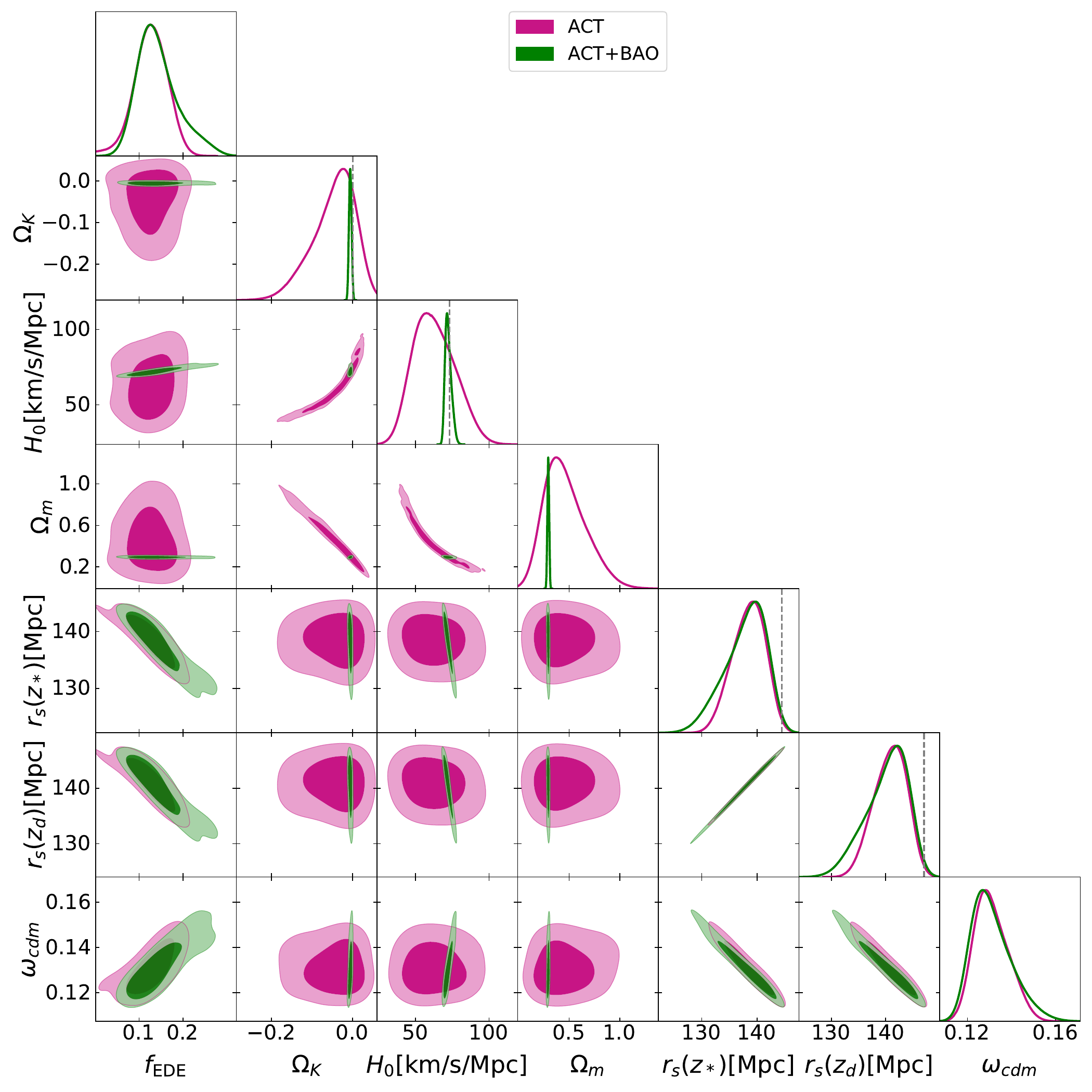}
    \caption{The same as Fig.~\ref{Figure:PlanckBAO} but with ACT only (magenta) and ACT+BAO (green). 
    }
    \label{Figure:ACTBAO}
\end{figure}
%=========================%

%=========================%
\begin{figure}[t]
    \centering
    \includegraphics [width=1\textwidth]{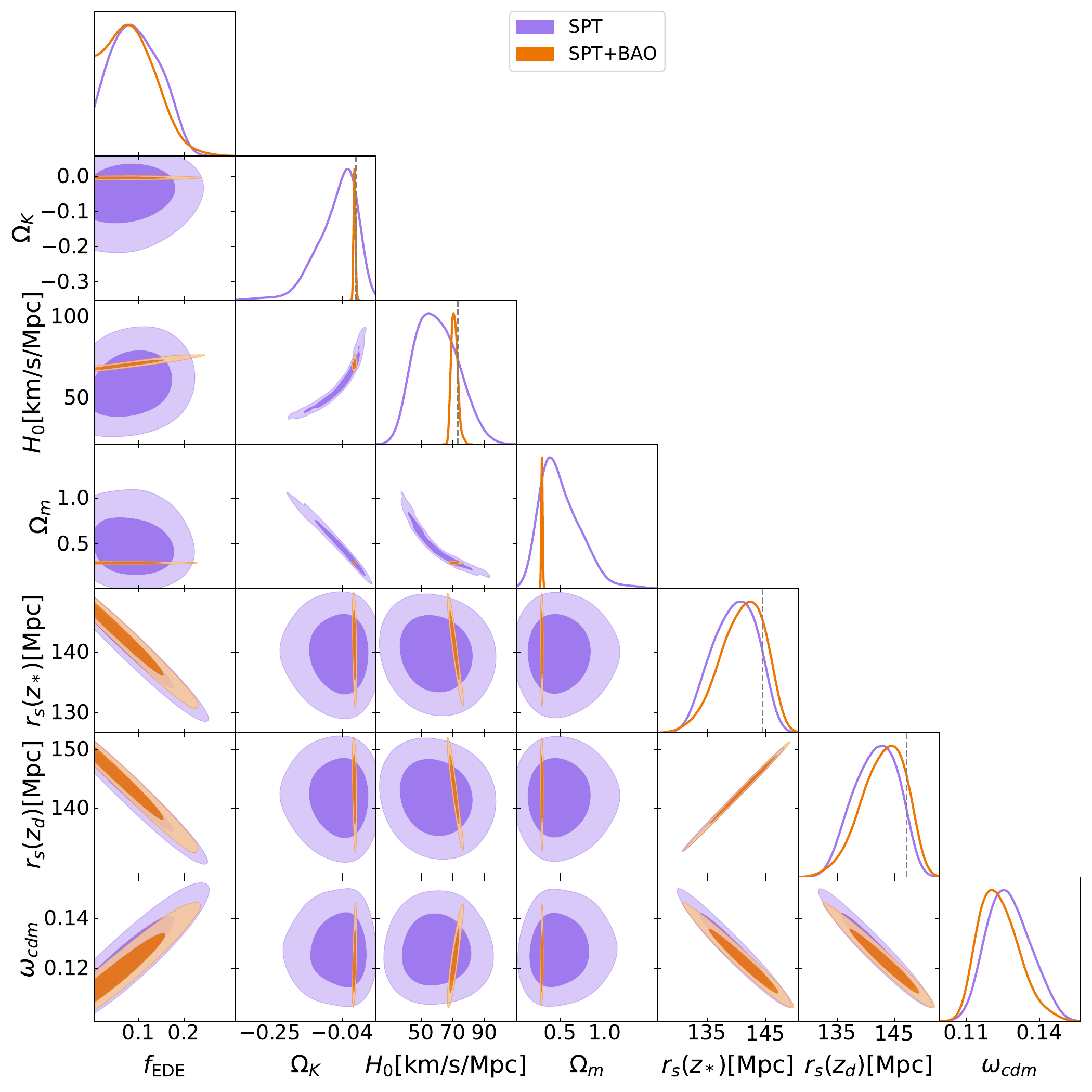}
    \caption{The same as Fig.~\ref{Figure:PlanckBAO} but with SPT only (purple) and SPT+BAO (orange). 
    }
    \label{Figure:SPTBAO}
\end{figure}
%=========================%

%%chi2 table:
\begin{table}[t]
\centering
\renewcommand{\arraystretch}{1.7} % Adjust the value to increase/decrease line spacing
\scalebox{0.55}{
\begin{tabular}{|l|c|c|c|c|}
\hline 
Dataset & $\Lambda \rm CDM$   & $\Lambda \rm CDM +\Omega_K+A_L$ & EDE & EDE$ \rm +\Omega_K+A_L$ \\
\hline
ACT DR4 (TT+TE+EE) & $287.3$ & $287.3$ &  $253.4$ & $274.7$\\
SPT-3G (TE+EE) & $1127.3$ & $1127.8$ & $1124.6$ & $1123.3$\\
Planck 2018 low$\ell$ TT (\textcolor{violet}{ACT}, \textcolor{orange}{SPT}, \textcolor{cyan}{Planck}) & $(\textcolor{violet}{21.7}, \textcolor{orange}{20.3}, \textcolor{cyan}{23.5})$ & $(\textcolor{violet}{21.5}, \textcolor{orange}{20.1}, \textcolor{cyan}{22.7})$ & $(\textcolor{violet}{21.5}, \textcolor{orange}{20.0}, \textcolor{cyan}{20.4})$ & $(\textcolor{violet}{21.4}, \textcolor{orange}{20.0}, \textcolor{cyan}{20.9})$ \\
Planck 2018 low$\ell$ EE (\textcolor{violet}{ACT}, \textcolor{orange}{SPT}, \textcolor{cyan}{Planck}) & $(\textcolor{violet}{396.4}, \textcolor{orange}{396.6}, \textcolor{cyan}{396.2})$ & $(\textcolor{violet}{396.3}, \textcolor{orange}{396.0}, \textcolor{cyan}{397.1})$ & $(\textcolor{violet}{396.4}, \textcolor{orange}{396.0}, \textcolor{cyan}{396.5})$ & $(\textcolor{violet}{396.5}, \textcolor{orange}{396.0}, \textcolor{cyan}{395.7})$ \\
Planck 2018 TT ($\ell_{\rm max}=650$), (\textcolor{violet}{ACT}, \textcolor{orange}{SPT}) &$(\textcolor{violet}{255.7}, \textcolor{orange}{254.5})$ & $(\textcolor{violet}{255.5}, \textcolor{orange}{254.3})$ & $(\textcolor{violet}{253.4}, \textcolor{orange}{254.5})$ & $(\textcolor{violet}{251.8}, \textcolor{orange}{255.3})$ \\
Planck 2018 high$\ell$ (TT+TE+EE) & $2350.2$ & $2347.1$ & $2350.4$ & $2342.0$ \\
Planck 2018 Lensing & $8.9$ & $9.5$ & $10.0$ & $9.6$ \\
\hline
BAO 6dF (\textcolor{violet}{ACT}, \textcolor{orange}{SPT}, \textcolor{cyan}{Planck}) & (\textcolor{violet}{$0.05$}, \textcolor{orange}{$0.07$}, \textcolor{cyan}{$0.05$}) & (\textcolor{violet}{$0.02$}, \textcolor{orange}{$0.07$}, \textcolor{cyan}{$0.03$}) & (\textcolor{violet}{$0.03$}, \textcolor{orange}{$0.04$}, \textcolor{cyan}{$0.06$}) & (\textcolor{violet}{$0.03$}, \textcolor{orange}{$0.03$}, \textcolor{cyan}{$0.03$})  \\
BAO MGS DR7 (\textcolor{violet}{ACT}, \textcolor{orange}{SPT}, \textcolor{cyan}{Planck}) & (\textcolor{violet}{$1.1$}, \textcolor{orange}{$2.2$}, \textcolor{cyan}{$1.3$}) & (\textcolor{violet}{$1.41$}, \textcolor{orange}{$2.3$}, \textcolor{cyan}{$1.8$})  & (\textcolor{violet}{$2.25$}, \textcolor{orange}{$2.3$}, \textcolor{cyan}{$1.5$}) & (\textcolor{violet}{$2.3$}, \textcolor{orange}{$1.8$}, \textcolor{cyan}{$1.5$})  \\
BAO BOSS DR12 (\textcolor{violet}{ACT}, \textcolor{orange}{SPT}, \textcolor{cyan}{Planck}) & (\textcolor{violet}{$4.3$}, \textcolor{orange}{$3.6$}, \textcolor{cyan}{$3.9$}) & (\textcolor{violet}{$4.02$}, \textcolor{orange}{$3.7$}, \textcolor{cyan}{$4.3$})  & (\textcolor{violet}{$3.5$}, \textcolor{orange}{$3.7$}, \textcolor{cyan}{$3.7$}) & (\textcolor{violet}{$4.02$}, \textcolor{orange}{$4.0$}, \textcolor{cyan}{$3.7$})  \\
BAO eBOSS DR14 Ly$\alpha$-auto (\textcolor{violet}{ACT}, \textcolor{orange}{SPT}, \textcolor{cyan}{Planck}) & (\textcolor{violet}{$1.7$}, \textcolor{orange}{$1.3$}, \textcolor{cyan}{$1.5$}) & (\textcolor{violet}{$1.4$}, \textcolor{orange}{$1.2$}, \textcolor{cyan}{$2.5$}) & (\textcolor{violet}{$1.3$}, \textcolor{orange}{$1.2$}, \textcolor{cyan}{$1.4$}) & (\textcolor{violet}{$1.1$}, \textcolor{orange}{$1.2$}, \textcolor{cyan}{$1.4$})  \\
BAO eBOSS DR14 Ly$\alpha$-cross (\textcolor{violet}{ACT}, \textcolor{orange}{SPT}, \textcolor{cyan}{Planck}) & (\textcolor{violet}{$4.7$}, \textcolor{orange}{$4.1$}, \textcolor{cyan}{$4.4$}) & (\textcolor{violet}{$4.3$}, \textcolor{orange}{$4.0$}, \textcolor{cyan}{$5.1$}) & (\textcolor{violet}{$4.2$}, \textcolor{orange}{$4.0$}, \textcolor{cyan}{$4.4$}) & (\textcolor{violet}{$3.7$}, \textcolor{orange}{$3.9$}, \textcolor{cyan}{$4.3$})  \\
BAO eBOSS DR14 Ly$\alpha$-combined (\textcolor{violet}{ACT}, \textcolor{orange}{SPT}, \textcolor{cyan}{Planck}) & (\textcolor{violet}{$5.1$}, \textcolor{orange}{$4.3$}, \textcolor{cyan}{$4.8$}) & (\textcolor{violet}{$4.7$}, \textcolor{orange}{$4.2$}, \textcolor{cyan}{$4.7$}) & (\textcolor{violet}{$4.4$}, \textcolor{orange}{$4.1$}, \textcolor{cyan}{$4.7$}) & (\textcolor{violet}{$3.7$}, \textcolor{orange}{$4.2$}, \textcolor{cyan}{$4.6$})  \\
BAO eBOSS DR16 Ly$\alpha \times $Ly$\alpha$ (\textcolor{violet}{ACT}, \textcolor{orange}{SPT}, \textcolor{cyan}{Planck}) & (\textcolor{violet}{$0.9$}, \textcolor{orange}{$0.7$}, \textcolor{cyan}{$0.8$}) & (\textcolor{violet}{$0.7$}, \textcolor{orange}{$0.6$}, \textcolor{cyan}{$1.2$}) & (\textcolor{violet}{$0.7$}, \textcolor{orange}{$0.6$}, \textcolor{cyan}{$0.8$}) & (\textcolor{violet}{$0.5$}, \textcolor{orange}{$0.6$}, \textcolor{cyan}{$0.7$})  \\
BAO eBOSS DR16 Ly$\alpha \times $QSO (\textcolor{violet}{ACT}, \textcolor{orange}{SPT}, \textcolor{cyan}{Planck}) & (\textcolor{violet}{$1.6$}, \textcolor{orange}{$1.3$}, \textcolor{cyan}{$1.5$}) & (\textcolor{violet}{$1.4$}, \textcolor{orange}{$1.3$}, \textcolor{cyan}{$1.8$}) & (\textcolor{violet}{$1.4$}, \textcolor{orange}{$1.3$}, \textcolor{cyan}{$1.5$}) & (\textcolor{violet}{$1.02$}, \textcolor{orange}{$1.2$}, \textcolor{cyan}{$1.4$})  \\
BAO eBOSS DR16 QSO (\textcolor{violet}{ACT}, \textcolor{orange}{SPT}, \textcolor{cyan}{Planck}) & (\textcolor{violet}{$0.6$}, \textcolor{orange}{$0.5$}, \textcolor{cyan}{$0.5$} ) & (\textcolor{violet}{$0.5$}, \textcolor{orange}{$0.5$}, \textcolor{cyan}{$0.3$}) & (\textcolor{violet}{$0.6$}, \textcolor{orange}{$0.4$}, \textcolor{cyan}{$0.5$}) & (\textcolor{violet}{$0.2$}, \textcolor{orange}{$0.2$}, \textcolor{cyan}{$0.5$})  \\
\hline
Total $\chi^2$ (CMB) (\textcolor{violet}{ACT}, \textcolor{orange}{SPT}, \textcolor{cyan}{Planck}) & (\textcolor{violet}{$961.3$}, \textcolor{orange}{$1798.8$}, \textcolor{cyan}{$2778.8$}) & (\textcolor{violet}{$960.8$}, \textcolor{orange}{$1798.6$}, \textcolor{cyan}{$2776.5$}) & (\textcolor{violet}{$945.02$}, \textcolor{orange}{$1795.4$}, \textcolor{cyan}{$2777.3$}) & (\textcolor{violet}{$944.8$}, \textcolor{orange}{$1795.0$}, \textcolor{cyan}{$2768.2$})  \\
Total $\chi^2$ (CMB+BAO) (\textcolor{violet}{ACT}, \textcolor{orange}{SPT}, \textcolor{cyan}{Planck}) & (\textcolor{violet}{$981.4
$}, \textcolor{orange}{$1822.4$}, \textcolor{cyan}{$2798.8$}) & (\textcolor{violet}{$979.9$}, \textcolor{orange}{$1816.8$}, \textcolor{cyan}{$2795.8$})  & (\textcolor{violet}{$968.5$}, \textcolor{orange}{$1817.5$}, \textcolor{cyan}{$2795.8$}) & (\textcolor{violet}{$962.3$}, \textcolor{orange}{$1817.1$}, \textcolor{cyan}{$2788.2$})  \\
$ \Delta \chi^2$/ (d.o.f) (CMB) (\textcolor{violet}{ACT}, \textcolor{orange}{SPT}, \textcolor{cyan}{Planck}) &  & (\textcolor{violet}{$-0.003$}, \textcolor{orange}{$-0.0001$}, \textcolor{cyan}{$-0.09$}) & (\textcolor{violet}{$-0.09$}, \textcolor{orange}{$-0.03$}, \textcolor{cyan}{$-0.02$}) & (\textcolor{violet}{$-0.09$}, \textcolor{orange}{$-0.03$}, \textcolor{cyan}{$-0.14$})  \\
$ \Delta \chi^2 $/ (d.o.f) (CMB+BAO) (\textcolor{violet}{ACT}, \textcolor{orange}{SPT}, \textcolor{cyan}{Planck}) &  & (\textcolor{violet}{$-0.008$}, \textcolor{orange}{$-0.04$}, \textcolor{cyan}{$-0.03$}) & (\textcolor{violet}{$-0.07$}, \textcolor{orange}{$-0.04$}, \textcolor{cyan}{$-0.04$}) & (\textcolor{violet}{$-0.1$}, \textcolor{orange}{$-0.04$}, \textcolor{cyan}{$-0.13$})  \\

\hline 
\end{tabular} 
}
\caption{Best-fit $\chi^2$ values for $\Lambda \rm CDM$, $\Lambda \rm CDM + \Omega_K + A_L$, $\rm EDE$ and $\rm EDE + \Omega_K + A_L $ models are presented in each column. Different combinations of CMB datasets (upper panel) are as follows: Full Planck: Planck-high-$\ell$ (\texttt{TT+TE+EE}) $+$ Planck-low-$\ell$ \texttt{TT} $+$ Planck-low-$\ell$ \texttt{EE} + Planck-lensing. ACT: ACT DR4 (\texttt{TT+TE+EE}) $+$ Planck-low-$\ell$ \texttt{TT} $+$ Planck-low-$\ell$ \texttt{EE} $+$ Planck \texttt{TT} ($\ell_{\rm max}=650$). SPT: SPT-3G (\texttt{TE+EE}) $+$ Planck-low-$\ell$ \texttt{TT} $+$ Planck-low-$\ell$ \texttt{EE} $+$ Planck \texttt{TT} ($\ell_{\rm max}=650$). The $\chi^2$ values represented in the upper panel show the best-fit values for only CMB runs while the middle panel values come from the (CMB+BAO) analysis. Note that the Gaussian prior on $\tau$ is applied in all analyses apart from the full Planck results. The $\chi^2$ best-fit values for some likelihoods that have been used for different data combinations such as ACT, SPT and Planck are shown in purple, orange and cyan respectively. The lower panel shows the total $\chi^2$ value for each CMB or CMB+BAO analysis. The $\Delta \chi^2$ values present the differences between the best-fit $\chi^2$ value and the $\Lambda \rm CDM$ model in each CMB and CMB+BAO case.  
}
\label{Table:chi2}
\end{table}

%=========================%
\begin{figure}[t]
    \centering
    \includegraphics [width=0.8\textwidth]{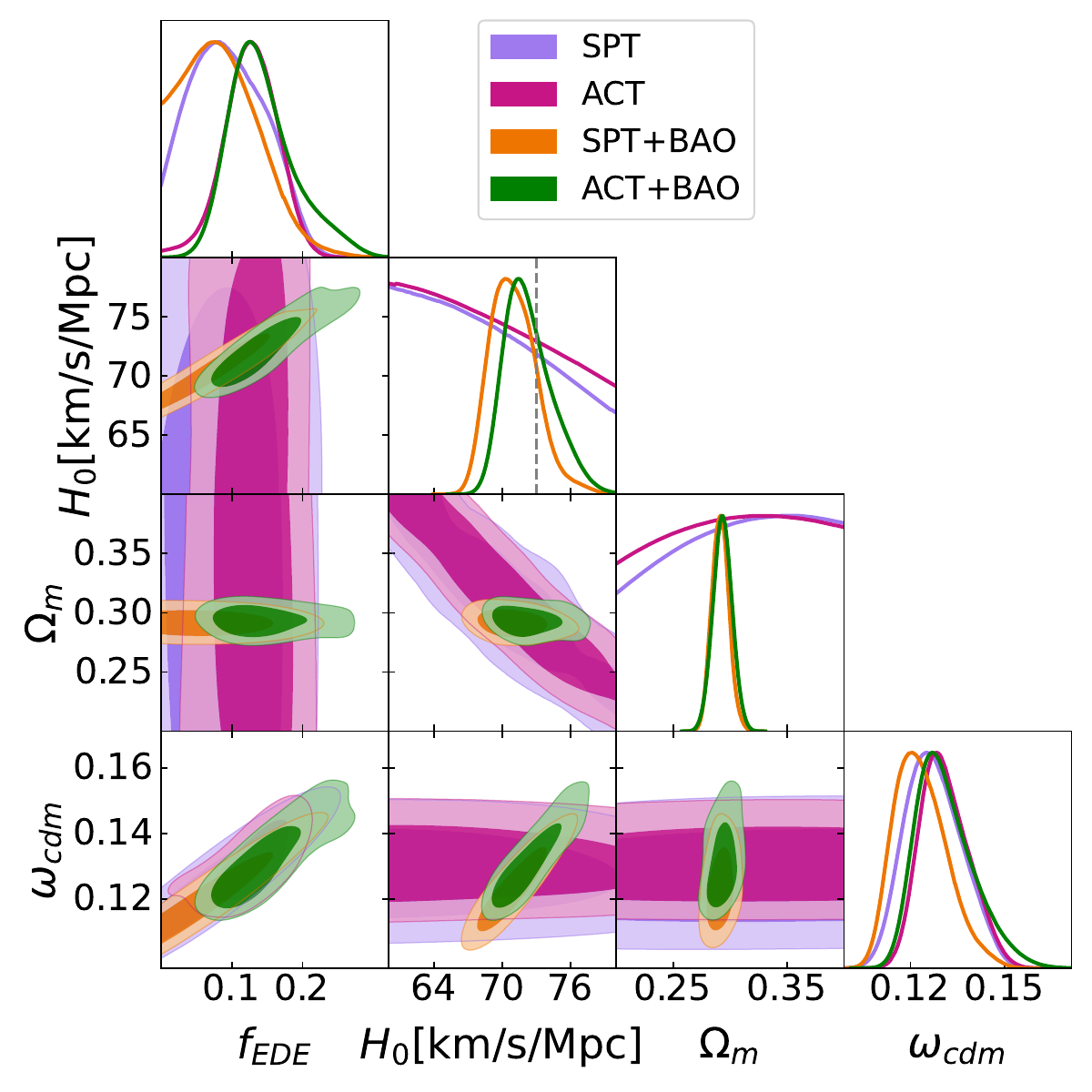}
    \caption{The same results from Figs.~\ref{Figure:ACTBAO} and \ref{Figure:SPTBAO} are combined for a clear comparison.  
    }
    \label{Figure:combinedBAO}
\end{figure}
%=========================%

Thus, it is crucial to combine the CMB data with other low-redshift probes to break the degeneracy and improve the CMB curvature constraints. 
In Figs.~\ref{Figure:PlanckBAO}, \ref{Figure:ACTBAO}, and \ref{Figure:SPTBAO}, we compare the CMB only result with the case combined with BAO for Planck, ACT, and SPT, respectively. 
Not surprisingly, adding BAO to the CMB data is extremely powerful to precisely measure $\Omega_{K}$; $\Omega_K = -0.0009_{-0.0019}^{+0.0019}$ for Planck+BAO, $\Omega_K = -0.0058_{-0.0031}^{+0.0031}$ for ACT+BAO, and $\Omega_K =-0.0032_{-0.0032}^{+0.0032}$ for SPT+BAO.
It is interesting to confirm that the current CMB+BAO data prefer $\Omega_{K}=0$ within 95\% C.L. even after marginalizing over the EDE parameters and hence the sound horizon scale. 
Since both the sound-horizon scales $r_{s}(z_{*})$ and $r_{s}(z_{d})$ are reduced by the existence of EDE by a similar amount, this result suggests that the paremeters important for the late-time universe such as $\Omega_{K}$ and $\Omega_{m}$ are mainly determined by the relative ratio of two standard rulers at two distinct epochs from the CMB and BAOs. 
To demonstrate this, we show in Fig.~\ref{Figure:spectra} the ratio of the sound horizon scale for EDE$+\Omega_K+A_L$ model with respect to the best-fit Planck $\Lambda$CDM. 
The middle panel shows the low-redshift BAO measurements, while the right panel shows the $\theta_s$ constraints of the same curved EDE model for Planck, ACT, and SPT measurements. 
The green curve is our best-fit curved EDE model for ACT, equivalent to the best-fit for the green contours in Fig.~\ref{Figure:ACTBAO}. 
The black dashed/dotted-dashed curves show the EDE$+\Omega_K+A_L$ model with slightly larger/smaller values for $\Omega_{K}$. 
Note that, since the $y$ axis is divided by the sound horizon, the reduction of the sound horizon is taken into account.  
The relative comparison between the CMB and BAO data points is crucial to break the degeneracies and therefore to precisely constrain $\Omega_K$. 

Table \ref{Table:chi2} shows the $\chi^2$ values for the best-fit $\Lambda \rm CDM$, $\Lambda \rm CDM + \Omega_K + A_L$, $\rm EDE$ and $\rm EDE + \Omega_K + A_L $ models with the contribution of each single likelihood. 
Looking at the $\Delta \chi^2/(\rm d.o.f)$ values at the bottom of the table, both flat-EDE and curved-EDE best-fit models, no matter which CMB data we use, shows an improvement over the best-fit $\Lambda \rm CDM$ model. This behavior remains the same when we include the BAO data. The $\rm EDE + \Omega_K + A_L $ model is slightly favored by the Planck CMB data rather than ACT or SPT. When we take BAO dataset into account the model still shows improvement against the $\Lambda \rm CDM$ model. 

Accurate determination of $\Omega_{K}$ and $\Omega_{m}$ with BAO data has a nonnegligible impact on other parameters by breaking degeneracies. 
$\Omega_{K}$ and $\Omega_{m}$ are correlated with $H_{0}$ through geometric degeneracy positively and negatively, respectively. 
Therefore, a higher $\Omega_{K}$ (or smaller $\Omega_{m}$) in CMB+BAO than CMB only generally leads to a higher $H_{0}$.
The $H_{0}$ value in Planck+BAO is not as high as the distance ladder measurement (the dashed vertical line), while those in ACT+BAO and SPT+BAO are more consistent with them. 
This result is driven by the preferred value in $f_{\rm EDE}$ and hence how much the sound horizon scale is reduced.
Interestingly, the ACT+BAO posterior has a peak at non-zero $f_{\rm EDE}$ consistently with the CMB-only case, while the SPT case has a notable (but statistically insignificant) shift in $f_{\rm EDE}$. 
To understand the reason behind this, we compare the ACT and SPT results in Fig.~\ref{Figure:combinedBAO}. 
The peak of $\Omega_{m}$ for the SPT data only is slightly deviated from the value preferred by BAO, resulting in a lower value of $\omega_{cdm}$.  
Since $\omega_{cdm}$ and $f_{\rm EDE}$ are positively correlated, the preferred value of $f_{\rm EDE}$ becomes smaller than the case without BAO. 
On the contrary, this is not the case for ACT, as the peak of $\Omega_{m}$ for ACT remains unchanged with or without the addition of BAO.

%%%%%%%%%%%%%%%%%%%%%%%%%%%%%%%%%%%%%%%%%%%%%%%%%%%%%%%%%%%%%%%%%%%%%%%%%%%%%%%%%%%%%
%%%%%%%%%%%%%%%%%%%%%%%%%%%%%%%%%%%%%%%%%%%%%%%%%%%%%%%%%%%%%%%%%%%%%%%%%%%%%%%%%%%%%

%%%%%%%%%%%%%%%%%%%%%%%%%%%%%%%%%%%%%%%%%%%%%%%%%%%%%%%%%%%%%%%%%%%%%%%%%%%%%%%%%%%%%
%%%%%%%%%%%%%%%%%%%%%%%%%%%%%%%%%%%%%%%%%%%%%%%%%%%%%%%%%%%%%%%%%%%%%%%%%%%%%%%%%%%%%
\section{Summary and discussion}
\label{sec:summary}
In this work, our main interest is to derive the constraint on the spatial curvature from the cosmic expansion history in a cosmological scenario with non-standard sound horizon scale. 
More specifically, we use recent CMB datasets from Planck, ACT, and SPT as well as the most updated BAO measurement from SDSS DR16 in both of which the change in the sound horizon scale would impact the inference on the spatial curvature parameter. 
As a working example, we adopt an axion-like EDE model which attracts attention in the community in light of the Hubble tension. 
We extend our EDE cosmological parameter space by the spatial curvature and the lensing amplitude to use the information from the cosmic expansion history. 
Our main findings are highlighted as follows:
\begin{itemize}
    \item We find that, independent of the CMB datasets, the EDE model parameters are constrained only by the CMB power spectra as precisely as the flat case in previous work, even with $\Omega_{K}$ and $A_{L}$. Although Planck disfavors non-zero $f_{\rm EDE}$, ACT and SPT alone prefer non-zero value, $f_{\rm EDE}< 0.218\,(0.199)$ at $95\%$ C.L. for ACT (SPT), (see Fig.~\ref{Figure:curvedcmb} and Table \ref{Table:main}). 
    Not surprisingly, the constraints on $\Omega_{K}$ and $H_{0}$ are weakly constrained by any CMB-only cases due to the geometric degeneracy.
    \item We demonstrate that combining CMB with BAO is extremely powerful to constrain the spatial curvature even with the reduction of the sound horizon scales. In the case of ACT which prefers the largest amount of $f_{\rm EDE}\sim 0.14$ which reduces the sound horizon scales by about 5\%, we obtain $\Omega_{K}=-0.0058\pm 0.0031$ after marginalizing over the EDE parameters. 
    This constraint is as competitive as the Planck + DR16 BAO result in a $\Lambda$CDM model, $\Omega_{K}=-0.0056\pm 0.0018$ \cite{Alam:2021eb} (see Figs.~\ref{Figure:OmKcomp}, \ref{Figure:spectra}, \ref{Figure:PlanckBAO}-\ref{Figure:SPTBAO} and Table \ref{Table:main}).  
\end{itemize}

Let us clarify the difference between our work and similar other work. 
Ref.~\cite{Fondi:2022tt} studied an EDE model that included $\Omega_{K}$ and $A_{L}$ using the Planck data and obtained $\Omega_K=-0.0007 \pm 0.0020$ when they combined with the SDSS DR12 BAO data. 
Although this is in excellent agreement with our result for a similar case, $\Omega_{K}=-0.0009_{-0.0019}^{+0.0019}$, there are a few minor differences between the two works. 
First, they consider Planck only, while we also study ACT and SPT. 
Second, their prediction of the EDE model is based on an approximated method implemented in \texttt{CAMB} to solve the linear perturbation equation for an EDE scalar field \cite{Smith:2019ih}. 
Instead, we adopt the exact method implemented in \texttt{CLASS} for an EDE model, following \cite{Hill:2020pl,Smith:2022aa}. 
In fact, we have tested the approximate method in \texttt{CAMB} and confirmed that we were unable to reproduce the previous work for ACT and SPT in \cite{Hill:2020pl,Smith:2022aa}. 
However, the impact of the approximation on the Planck case is minor, since Planck does not prefer nonzero $f_{\rm EDE}$. 
Finally, they vary the equation-of-state parameter for the late-time dark energy, $w$, while we fix it with the cosmological constant, $w=-1$. 
As Ref.~\cite{Fondi:2022tt} showed in their Fig.~2, a degeneracy between $\Omega_{K}$ and $w$ is expected to some extent, but their Planck+BAO result is in perfect agreement with $w=-1$. 
Along the similar line, there are similar studies which consider different dark energy scenarios which become prominent \textit{well after} the recombination (see e.g., \cite{Ryan:2019mn,Khadka:2020mn,Khoraminezhad:2020cer,Cao:2022mn,Cruz:2022oqk,Yang:2022ex,deCruzPerez:2022ar,Reboucas:2023aa,Haridasu:2022dyp,Corona:2021qxl}). 
We leave the impact of these other extended parameters on the spatial curvature for future work. 

As motivated by Fig.~\ref{Figure:spectra}, it is desirable to obtain more accurate and precise data from CMB such as LiteBird, CMB-S4, and the Simons Array (see e.g., \cite{Chang:2022sn} for a recent review) as well as the BAO data filling at $z\gtrsim 1$ such as Hobby-Eberly Telescope Dark Energy Experiment \cite{Gebhardt:2021mm}, Dark Energy Spectroscopic Instrument \cite{DESI:2016pp}, Prime Focus Spectrograph \cite{Takada:2014aa}, and Roman Space Telescope \cite{Wang:2022tt}.
These new datasets will lead to a more precise and accurate constraint on the spatial curvature of the Universe (see e.g., \cite{Takada:2015ok,Leonard:2016ph,Sailer:2021aa,Moradinezhad:2023ar}).

%%%%%%%%%%%%%%%%%%%%%%%%%%%%%%%%%%%%%%%%%%%%%%%%%%%%%%%%%%%%%%%%%%%%%%%%%%%%%%%%%%%%%
%%%%%%%%%%%%%%%%%%%%%%%%%%%%%%%%%%%%%%%%%%%%%%%%%%%%%%%%%%%%%%%%%%%%%%%%%%%%%%%%%%%%%
%\appendix
%\section{Some title}
%Please always give a title also for appendices.

\acknowledgments
We would like to thank anonymous referee for his/her constructive comments. JS was supported by the OURE program at Missouri University of Science and Technology. 
HK and SS acknowledge the support for this work from NSF-2219212.
SS is supported in part by World Premier International Research Center Initiative (WPI Initiative), MEXT, Japan.

\appendix
\section{Planck results with fixed $A_{L}=1$}
\label{sec:appendix}

\begin{figure}[t]
    \centering
    \includegraphics [width=0.8\textwidth]{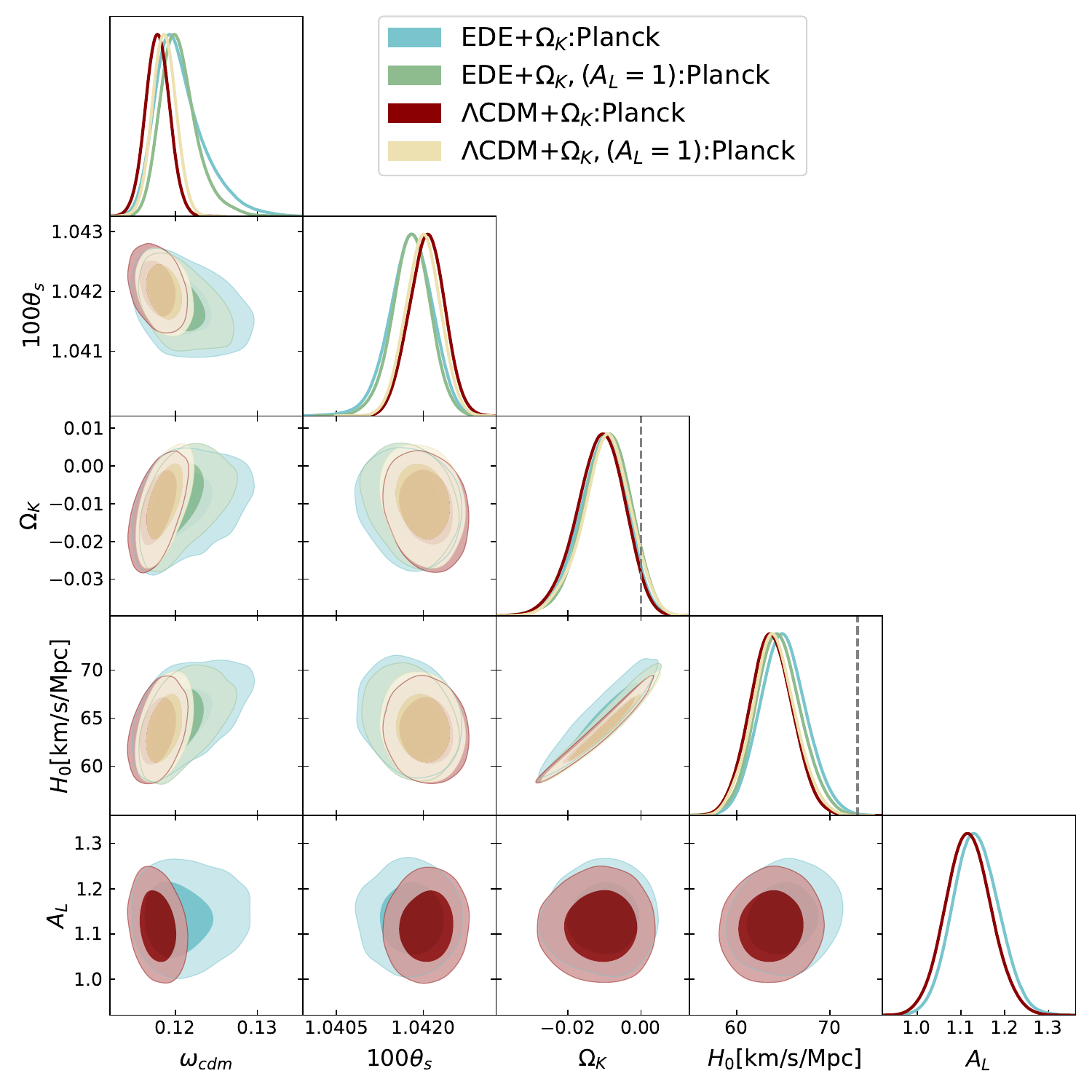}
    \caption{1D and 2D posterior distributions (68\% and 95\% C.L. for the 2D contours) of various cosmological parameters in a curved universe for both $\Lambda$CDM and EDE models using Planck CMB dataset. Green and yellow contours demonstrate the results for $A_L=1$, in EDE and $\Lambda$\rm CDM models respectively.} 
    \label{Figure:app1}
\end{figure}

\begin{figure}[t]
    \centering
    \includegraphics [width=0.8\textwidth]{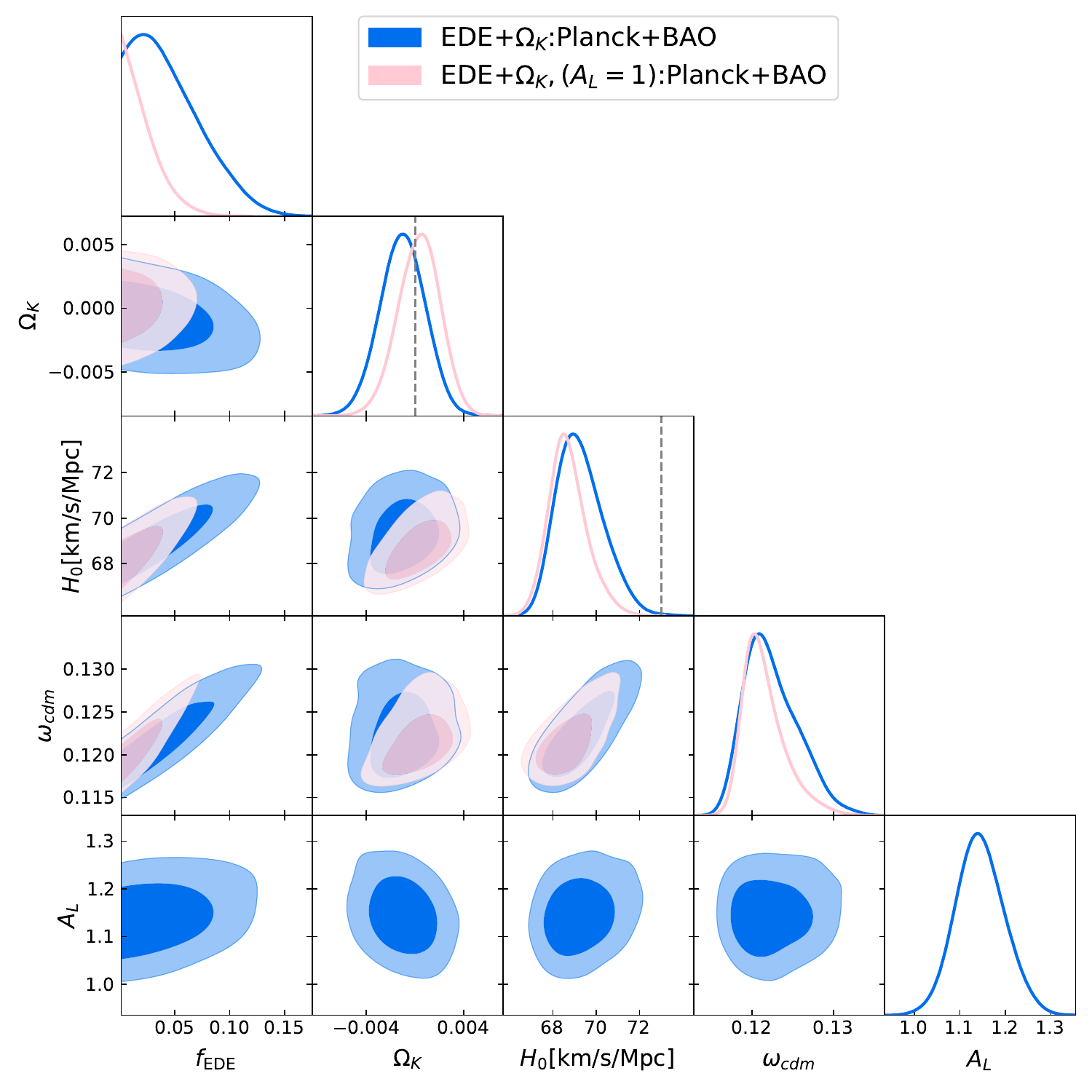}
    \caption{Posterior distributions for some key parameters in the curved EDE model taking the BAO dataset into account. The blue contours show the curved EDE scenario with varying the $A_L$ parameter while pink contours represent the same scenario with fixed $A_L=1$.}
    \label{Figure:app2}
\end{figure}

In this appendix, we present our results for the full Planck dataset and we compare the posteriors with the fixed and varied $A_{L}$ parameter. 
Our focus in this work was mainly on how the spatial curvature is constrained by distance measurements from the sound horizon scale in CMB and BAO measurements when the sound horizon is reduced by an EDE model. 
By varying the $A_L$ parameter, we marginalized the information from the CMB lensing. 
Nonetheless, since Planck TT prefers the $A_{L}$ value to be larger than one, we show the Planck results when we fix it to $A_L=1$. 
Fig~\ref{Figure:app1}, shows a comparison between EDE and $\Lambda$CDM curved model posteriors with and without fixing $A_L$ parameter. 
The cyan and brown contours are the EDE and $\Lambda$CDM models when the $A_L$ parameter is being varied (equivalent to the same color contours in Fig~\ref{Figure:curvedcmb}) while in green and yellow contours we fix $A_L=1$. 
The constrains on $\Omega_K$ parameter in the case of EDE model changes from $\Omega_K = -0.0105 \pm 0.0066 $ (for varied $A_L$) to $\Omega_K = -0.0098 \pm 0.0068 $ (for fixed $A_L$). 
The $H_0$ parameter is constrained as $H_0 = 64.93 \pm 2.57 $ (for varied $A_L$) and $H_0 = 64.37 \pm 2.50 $ (for fixed $A_L$). The constraint on $f_{\rm EDE}$ parameter would changes from $f_{\rm EDE} = 0.033 \pm 0.026$ (in the case of varied $A_L$) to $f_{\rm EDE} =0.016 \pm 0.013$ (fixed $A_L$).
We also present a comparison between EDE$+\Omega_K + A_L$ and EDE$+\Omega_K$ models in light of the full Planck + BAO dataset in Fig~\ref{Figure:app2}. 
These results show that fixing the value of $A_L$ parameter lowers the mean value of the $f_{\rm EDE}$ parameter and reduces errors from $f_{\rm EDE} = 0.044 \pm 0.030$ to $f_{\rm EDE} = 0.022 \pm 0.017$ while increasing the mean value of the $\Omega_K$ parameters and reducing errors from $\Omega_K = -0.0009 \pm 0.0019$ to $\Omega_K = 0.0003 \pm 0.0018$.

%=========================%
%=========================%

%=========================%
%=========================%

\bibliography{ms}
\bibliographystyle{JHEP}

\end{document}